\documentclass[onecolumn, a4size, 11pt]{IEEEtran}

\usepackage{amsmath,amsfonts,amssymb}
\usepackage{graphicx}
\usepackage{cite}
\usepackage{multirow}
\usepackage{algorithm}
\usepackage{algorithmic}
\usepackage{float}
\usepackage{caption}
\usepackage{subcaption}
\usepackage{multirow}
\usepackage{url}
\usepackage{color}
\usepackage{array}

\graphicspath{{./figs/}}

\IEEEoverridecommandlockouts

\newtheorem{theorem}{Theorem}

\newtheorem{remark}{Remark}

\newcommand{\dofbar}{\overline{d}}

\newcommand{\xtd}{\tilde{\mathbf x}}
\newcommand{\ytd}{\tilde{\mathbf y}}
\newcommand{\std}{\tilde{\mathbf s}}
\newcommand{\Htd}{\tilde{\mathbf H}^{(L)}}

\newcommand{\HtdH}{\tilde{\mathbf H}^{(H)}}
\newcommand{\smallterm}{n \mathrm{o}(\log \rho)+\mathrm{o}(n)}
\newcommand{\Mtd}{\tilde M}
\newcommand{\Ntd}{\tilde N}

\begin{document}
\title{Degrees of Freedom of the $3$-User Rank-Deficient MIMO Interference Channel}
\author{Yong~Zeng,  Xiaoli~Xu, Yong~Liang~Guan, Erry~Gunawan, and Chenwei~Wang
\thanks{This work has been presented in part in IEEE International Conference on Communications (ICC), Budapest, Hungary, June 2013.}
\thanks{Y.~Zeng was with the School of Electrical and Electronic Engineering, Nanyang Technological University, Singapore. He is now with the Department of Electrical and Computer Engineering, National University of Singapore (email: elezeng@nus.edu.sg).}
\thanks{X. Xu, Y. L. Guan, and E. Gunawan are with the School of Electrical and Electronic Engineering, Nanyang Technological University, Singapore 639801 (email: \{xuxiaoli, eylguan, egunawan\}@ntu.edu.sg).}
\thanks{C. Wang is with the Mobile Network Technology Group, Docomo Innovations Inc., Palo Alto, CA 94304, USA (e-mail: chenweiw@uci.edu).}
}


\maketitle

\begin{abstract}
We provide the degrees of freedom (DoF) characterization for the $3$-user $M_T\times M_R$ multiple-input multiple-output (MIMO) interference channel (IC) with \emph{rank-deficient} channel matrices, where each transmitter is equipped with $M_T$ antennas and each receiver with $M_R$ antennas, and the interfering channel matrices from each transmitter to the other two receivers are of ranks $D_1$ and $D_2$, respectively. One important intermediate step for both the converse and achievability arguments is to convert the fully-connected rank-deficient channel into an equivalent partially-connected full-rank MIMO-IC by invertible linear transformations. As such, existing techniques developed for full-rank MIMO-IC can be incorporated to derive the DoF outer and inner bounds for the rank-deficient case.  Our result shows that when the interfering links are weak in terms of the channel ranks,  i.e., $D_1+D_2\leq \min(M_T, M_R)$, zero forcing is sufficient to achieve the optimal DoF. On the other hand, when $D_1+D_2> \min(M_T, M_R)$, a combination of zero forcing and interference alignment is  in general required for DoF optimality. The DoF characterization obtained in this paper unifies several existing results in the literature.


\end{abstract}

\section{Introduction}
  Finding the information-theoretic capacity of the  general Gaussian interference channel (IC) is a long-standing  open problem \cite{163}.  On the other hand, the characterizations of the degrees of freedom (DoF) value, a parameter that provides the first-order capacity approximation at the asymptotically high signal-to-noise ratio (SNR), have been successfully accomplished in various settings \cite{92,93,85,363,189,348,364,367,362}. In particular, for interference channels with more than two users  and hence each receiver is interfered by more than one interfering sources, the technique known as interference alignment has been shown to achieve higher DoF than previously believed \cite{85,363,189,348}. More recently, for the $3$-user $M_T \times M_R$ IC with $M_T$ antennas at each transmitter and $M_R$ antennas at each receiver, a novel concept of ``subspace alignment chains'' was introduced in \cite{362}, where ideally each transmitted signal subspace is aligned with another interference subspace at both undesired receivers and an alignment chain is formed, and the finite length of the alignment chain ultimately limits the attained DoF value. Furthermore, for the DoF outer bound proof in  \cite{362}, a technique termed ``onion peeling'' has been introduced, where an equivalent channel with layered structure was obtained by applying change of basis operations at the transmitters and receivers. As such, it helps determining the appropriate genie signals to be provided to each receiver  for deriving the outer bounds.



 The aforementioned works for DoF characterizations have all assumed full-rank channel matrices. Such an assumption is  valid in rich scattering environment with sufficient number of signal paths. However, in some practical scenarios, the communication channel matrices may be rank deficient due to poor scattering and hence with only very few signal paths. 
 There are only a few works reported on the DoF characterizations of rank-deficient multiple-input multiple-output IC (MIMO-IC) \cite{Chae11,Krishnamurthy12,492}.  In \cite{Chae11}, under the assumption that all channel matrices have the same rank, an achievable DoF value was obtained for the $K$-user time-varying  IC. The achievable scheme proposed therein is based on either zero forcing or interference alignment, whichever gives a higher DoF value. In \cite{Krishnamurthy12}, the authors investigated the DoF of the $2$-user and $3$-user rank-deficient MIMO-IC. While the optimal DoF value has been obtained for the general 2-user scenario, only the symmetric $M\times M$ setup where all terminals are equipped with $M$ antennas was considered for the 3-user case. Later, the result in \cite{Krishnamurthy12} was extended to the symmetric $K$-user $M\times M$ MIMO-IC \cite{492}, where all direct channels have rank $D_0$ and all cross channels have rank $D$. 

In this paper, we investigate the DoF  of the $3$-user $M_T \times M_R$ rank-deficient MIMO-IC where each transmitter is equipped with $M_T$ antennas and each receiver with $M_R$ antennas. All the direct channel matrices are of rank $D_0$, and the interfering channel matrices from each transmitter to the other two receivers are of ranks $D_1$ and $D_2$, respectively. Our model is more general than that studied in \cite{Krishnamurthy12} since the transmitters and receivers may have different number of antennas $M_T$ and $M_R$, respectively. It is worth mentioning that such a generalization is nontrivial due to the more complicated relations between the channel ranks and the number of transmit and receive antennas, which makes the derivations of the outer and inner bounds much more challenging. In fact, even for the full-rank case, the extension of the DoF result for the 3-user MIMO-IC from the $M\times M$ setup \cite{85} to the more general $M_T\times M_R$ case \cite{362} is highly involved and requires much more sophisticated techniques such as subspace alignment chains and onion peeling, as discussed previously. 
By adopting spatial extension for the inner bound to deal with non-integer DoF values \cite{362}, we provide a complete DoF characterization for all possible settings of the  $3$-user rank-deficient MIMO-IC parameterized by $(M_T, M_R, D_0, D_1, D_2)$. Specifically, with spatial extension, the DoF achievability is firstly established for a larger network obtained by virtually scaling the number of antennas and channel ranks so that the DoF  are integers. By normalizing with the same factor, the desired result for the original setup is then obtained. As argued in \cite{362} and \cite{365}, such a technique is justified by the fact that for every wireless network where the DoF characterizations are available, the DoF result is unaffected by spatial extension and then DoF normalization. Therefore, it has been conjectured that similar to time/frequency dimension, the DoF of a wireless network scale with the proportional scaling of spatial dimension \cite{362}. On the other hand, compared to time or frequency extensions, spatial extension makes it easier to deal with non-integer DoF values without having to deal with the added complexity of diagonal or block diagonal channel structures. 


To achieve the maximum possible DoF value, intuitively, one should zero force as many interfering links as possible. One distinguishing feature of the rank-deficient MIMO-IC, as compared to the full-rank counterparts, is that \emph{both} the left and right null space of the interfering channel matrices are non-empty, and hence should be utilized for zero forcing purposes.  Therefore, a critical step for achieving the optimal DoF for rank-deficient MIMO-IC is to appropriately exploit the rank deficiency of the interfering channel matrices. To this end, we first transform the original fully-connected rank-deficient MIMO-IC into an equivalent partially-connected \emph{full-rank} MIMO-IC. This in effect extracts the rank-deficiency of the interfering channel matrices embedded in all antennas to a subset of antennas so that the corresponding interfering links are nullified. As such, existing techniques developed for full-rank MIMO-IC, such as ``onion peeling'' based on the equivalent layered channel structure and subspace alignment chains, can be incorporated to facilitate the derivations of the DoF outer and inner bounds.


The  outer bound is derived with the genie-aided signaling technique based on the developed equivalent layered channel model. For the achievability, a two-layered linear processing scheme is proposed, with zero forcing in the inner layer and a \emph{block-diagonal} precoding  and interference cancelation, if necessary, in the outer layer. The use of block-diagonal precoding, together with the zero forcing operation in the inner layer,   ensures that each information-bearing symbol only interferes a particular group of receive antennas at the undesired receivers, which in turn makes interference cancelation possible. 
When the   interfering links are weak in terms of the channel ranks, i.e., $D_1+D_2\leq \min(M_T, M_R)$,  zero forcing together with a random block-diagonal precoding is sufficient to achieve the optimal DoF; whereas when $D_1+D_2>\min(M_T,M_R)$, the block-diagonal precoding in the outer layer must be carefully designed so that interference alignment is achieved, i.e., in this case, a  combination of zero forcing and interference alignment is  in general required to achieve the optimal DoF value.



\emph{Notations:} Scalars are denoted by italic letters. Boldface lower- and upper-case letters denote vectors and matrices, respectively. $\mathbb{C}^{N\times M}$  represents the space of $N\times M$ complex matrices. For a matrix $\mathbf A$, $\mathbf A^T$ and $\mathrm{rank}(\mathbf{A})$ denote its transpose and rank, respectively. The  null space of $\mathbf A$ is denoted by  $\mathcal{N}(\mathbf A)$, and hence its left null space can be represented as $\mathcal{N}(\mathbf A^T)$. $\mathrm{span}(\mathbf A)$ denotes a subspace spanned by the columns of $\mathbf A$. 
 $|\mathbf d|$ denotes the number of elements in vector $\mathbf d$.
For a real number $x$, $\lceil x \rceil$ denotes the smallest integer not less than $x$ and $\lfloor x \rfloor$ is the largest integer not greater than $x$. 
$\mathrm{o}(x)$ represents any function $f(x)$ such that $\lim_{x\rightarrow \infty} f(x)/x=0$.  

\section{System Model}\label{sec:systemModel}
We consider a $3$-user MIMO-IC where transmitter $k$ is intended to send message $W_k$ to receiver $k$, $k=1,2,3$.  Each transmitter is equipped with $M_T$ antennas and each receiver has $M_R$ antennas.  The signal received at receiver $k$ is given by
\begin{equation}\label{eq:channelModel}
\begin{aligned}
\mathbf{y}_k&=\mathbf{H}_{kk}\mathbf{x}_k+\sum_{i\neq k}\mathbf{H}_{ki}\mathbf{x}_i+\mathbf{n}_k, \ k=1,2,3,
\end{aligned}
\end{equation}
where $\mathbf{H}_{kk}\in\mathbb{C}^{M_R\times M_T}$ denotes the direct channel matrix from transmitter $k$ to receiver $k$, while $\mathbf{H}_{ki}\in\mathbb{C}^{M_R\times M_T}$, $i\neq k$, denotes the interfering channel matrix from transmitter $i$ to receiver $k$; $\mathbf{y}_k\in \mathbb{C}^{M_R\times 1}$ represents the received signal vector at receiver $k$; $\mathbf{x}_k\in \mathbb{C}^{M_T\times 1}$ is the transmitted signal vector  from transmitter $k$; and $\mathbf{n}_k\in \mathbb{C}^{M_R\times 1}$ denotes the additive  Gaussian noise vector.

\begin{figure}[htb]
\centering
\includegraphics[scale=0.55]{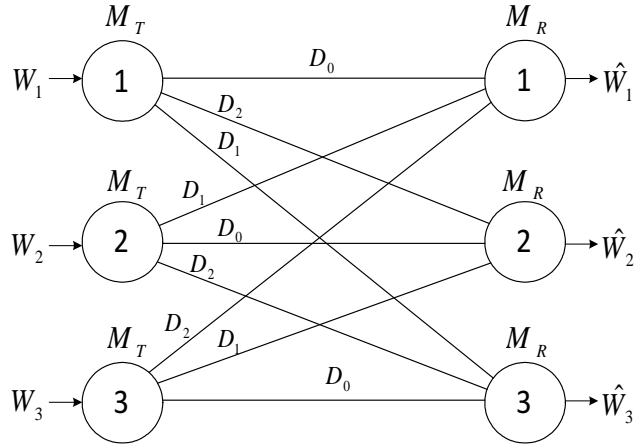}
\caption{$3$-user rank-deficient MIMO-IC parameterized by $(M_T, M_R, D_0, D_1, D_2)$.\vspace{-3ex}}
\label{F:3IC}
\end{figure}

 Different from most existing works where full-rank channel matrices have been assumed, here we consider the more general scenario with  possibly \emph{rank-deficient} channel matrices, which may occur in the poor scattering communication environment with very few signal paths as illustrated in Section~\ref{sec:numerical}. Specifically, all the direct channel matrices $\mathbf{H}_{kk}$ are assumed to be of rank $D_0$, and the interfering channel matrices $\mathbf{H}_{(k-1)k}$ and $\mathbf{H}_{(k+1)k}$ are of ranks $D_1$ and $D_2$, respectively, as shown in Fig.~\ref{F:3IC}. We then have $D_i\leq \min(M_R, M_T),\  i=0,1, 2$. Note that the user index $k$ is interpreted modulo $3$ so that, e.g., user $0$ is the same as user $3$.  Due to small-scale fading, the channel matrices are assumed to be independent and randomly generated, where a random rank-deficient channel matrix of size $M_R \times M_T$ with rank $D$ can be seen as a product of two full-rank generic matrices of size $M_R\times D$ and $D\times M_T$. In the following, this channel model is expressed compactly as $(M_T, M_R, D_0, D_1, D_2)$.

Denote by $\rho$ the average power constraint at each transmitter. The sum rate $R_{\Sigma}(\rho)$ and the DoF  $d_{\Sigma}$ follow from the standard definition in information theory. Further denote by $d=d_{\Sigma}/3$ the normalized DoF per user. Similar as \cite{362}, we define the \emph{spatially-normalized} sum DoF as
\begin{align}
\dofbar_{\Sigma}=\max_{q\in \mathbb{Z}^+}\frac{d_{\Sigma}(qM_T,qM_R,qD_0,qD_1,qD_2)}{q}.\label{E:dofbar}
\end{align}
Note that in \eqref{E:dofbar}, both the number of transmit and receive antennas and the ranks of the channel matrices are scaled by a factor $q$, and the DoF are normalized by the same factor.  While such a spatial extension is not necessary for the DoF outer bound, it is quite useful for the achievability result to deal with non-integer DoF values. 
 We further denote by $\dofbar$ the spatially-normalized DoF per user, i.e., $\dofbar=\dofbar_{\Sigma}/3$.
For notational convenience, in the remaining part of this paper, we let $D_t\triangleq D_1+D_2$, $N\triangleq\max\left(M_T, M_R\right)$, and $M\triangleq\min\left(M_T,M_R\right)$.

\section{Main Result and Discussions}\label{sec:mainResults}

\begin{theorem} \label{theo:main} 
For the $3$-user rank-deficient MIMO-IC $(M_T, M_R, D_0, D_1, D_2)$, the spatially-normalized DoF value per user is given by
\begin{align}\label{eq:exactDoF}
\dofbar=
\begin{cases}
\min\left(D_0,\frac{N+M-D_t}{2}\right), & D_t\leq M \\
\min\left(D_0,\frac{N}{2},\frac{pM}{2p-1},\frac{pN+2M-D_t}{2p+1}\right), & D_t>M
\end{cases},
\end{align}
where $p=\big\lceil\frac{D_t-M}{N-M}\big\rceil$.
\end{theorem}

The converse proof of Theorem~\ref{theo:main} is based on the genie-aided signaling technique. 
 The details are given in Section~\ref{sec:outerBound} with a preliminary step presented in Section~\ref{sec:changeOfBasis}. Note that the DoF converse is established for arbitrary values of $M_T$, $M_R$, $D_0$, $D_1$ and $D_2$, without the need for spatial extension. However, it can be verified that if the number of antennas and the ranks of the channel matrices are scaled by a factor $q$, the outer bound for the DoF value would be scaled by the same factor as well. Therefore, the outer bound for the rank-deficient MIMO-IC scales with spatial dimension, and hence holds both with and without spatial normalization.

The achievability proof of Theorem~\ref{theo:main} is based on a two-layered linear processing scheme, with zero forcing in the inner layer and block-diagonal precoding and interference cancelation in the outer layer.  The details are given in Section~\ref{sec:innerBound}.

\begin{figure} 
\centering
\includegraphics[scale=0.45]{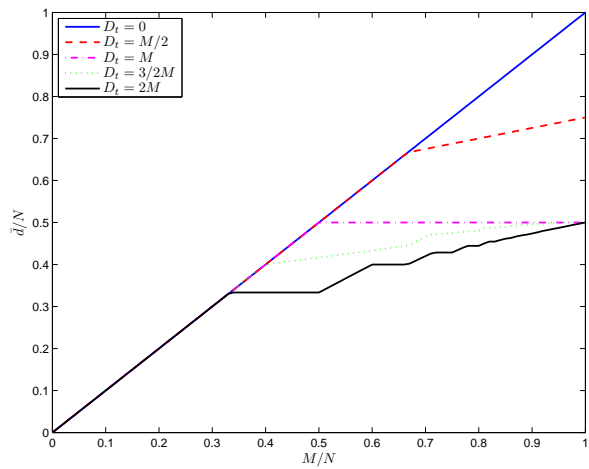}
\caption{$\bar d/N$ as a function of $M/N$ for different $D_t$ values, with $D_0=M$.\vspace{-2ex}}
\label{F:DoFPlot}
\end{figure}

\begin{remark}
Due to symmetry, the DoF value depends only on the \emph{sum} of the ranks of the interfering channel matrices $D_t=D_1+D_2$, instead of the individual ranks $D_1$ and $D_2$. Therefore, a network parameterized by $(M_T, M_R, D_0, D_1, D_2)$ would have identical spatially-normalized DoF value as that parameterized by $(M_T, M_R, D_0, D_1-\Delta_D, D_2+\Delta_D)$, for any valid $\Delta_D$.
\end{remark}
\begin{remark}
For the rank-deficient MIMO-IC with fixed number of antennas, the DoF value is determined  by either the rank of the direct link or that of the interfering links, depending on which one is the bottleneck. While the direct link affects the DoF via the simple quantity $D_0$, the effect of the interfering links is complicated by the relations between $D_t$, $M$ and $N$. With $D_0=M$, Fig.~\ref{F:DoFPlot} plots $\bar d/N$ versus $M/N$ for different $D_t$ values. It is observed that in the sufficiently low $M/N$ regime where either the transmitters (when $M_T>M_R$) or the receivers (when $M_R>M_T$) have enough number of antennas to zero force all interfering links, $\bar d$ is solely limited by the rank of the direct links, i.e., $\bar d=D_0=M$; otherwise, the interfering links will become the bottleneck and limit the maximum achievable $\bar d$. It is also observed for all $D_t\geq M$, the DoF converges to the ``half-cake'' result \cite{85} as $M$ approaches $N$. On the other hand, with $D_t<M$, higher DoF can be achieved, which is expected.
\end{remark}

The DoF characterization given in Theorem~\ref{theo:main} unifies several existing results in the literature. 

\subsubsection{$D_t=0$}
In the extreme case when the interfering links vanish, i.e., $D_1=D_2=0$, the model in \eqref{eq:channelModel} reduces to the $3$-user independent point-to-point MIMO channels with the direct channel matrices having rank $D_0$. In this case, it is well known that the DoF value per user is $D_0$ \cite{36}, which is equal to the value obtained by evaluating \eqref{eq:exactDoF} with $D_t=0$.

\subsubsection{$M_T=M_R$}
When restricting to the case where all terminals are equipped with the same number of antennas, i.e., $M_T=M_R$ or $M=N$,  the spatially-normalized DoF value given in Theorem~\ref{theo:main} reduces to that obtained in \cite{Krishnamurthy12}, which is
\begin{align}\label{eq:MequalN}
\dofbar=\min\left\{D_0,\max\left(\frac{M}{2},M-\frac{D_t}{2}\right)\right\}.
\end{align}


\subsubsection{$D_0=D_1=D_2=M$}
 For the setup when all channel matrices have full rank $M$, the spatially-normalized DoF have been obtained in  \cite{362}. By substituting $D_t$ with $2M$, the result given in Theorem~\ref{theo:main} can be simplified to
\begin{equation}\label{eq:fullrank}
\dofbar=\min\left\{\frac{pM}{2p-1},\frac{pN}{2p+1}\right\}, \textnormal{ where } p=\Big\lceil\frac{M}{N-M}\Big\rceil,
\end{equation}
which matches with that obtained in \cite{362}.

\section{Equivalent Partially-Connected Full-Rank  MIMO-IC}\label{sec:changeOfBasis}
 In this section, we present a preliminary step for the proof of Theorem~\ref{theo:main}, i.e., the conversion of the original fully-connected rank-deficient  MIMO-IC to an equivalent partially-connected \emph{full-rank}  MIMO-IC by exploiting the rank deficiency of the interfering links and applying \emph{invertible} linear transformations at the transmitters and receivers. While \emph{invertible} linear transformations, or change of basis operations, do not affect the DoF, they create an equivalent channel such that the existing techniques developed for the full-rank MIMO-IC can be incorporated to facilitate the DoF study of the  rank-deficient case.

Due to the reciprocity of linear transformations \cite{55}, without loss of generality, we assume in this section and in Section~\ref{sec:innerBound} that $M_T\leq M_R$, so that $M=M_T$ and $N=M_R$. Furthermore,  since different linear transformations are required for different interference levels, we distinguish the low- and high-interference cases, where the interference level is measured by the ranks of the interfering channel matrices as far as the DoF are concerned. Specifically, we have
\begin{enumerate}
\item{Low-interference case: $0\leq D_t\leq M$.}
\item{High-interference case: $M<D_t\leq 2M$.}
\end{enumerate}


\subsection{Low-Interference Case: $D_t\leq M$}\label{sec:CoBLow}
For the low-interference case with $D_t\leq M$, we start with the design of the invertible linear transformation $\mathbf R_k\in \mathbb{C}^{N\times N}$ at receiver $k$, which is based on the interfering channel matrices $\mathbf H_{k(k+1)}$ from transmitter $k+1$  and $\mathbf H_{k(k-1)}$ from transmitter $k-1$. 
 Based on which interfering links to be zero-forced, $\mathbf R_k$ is partitioned  as
\begin{equation}\label{eq:beamformer3}
\mathbf{R}_{k}=\left[\begin{matrix}\mathbf{U}_{k(k+1)} \\ \mathbf{U}_{k}^{c} \\ \mathbf{U}_{k(k-1)}\end{matrix}\right],
\end{equation}
where $\mathbf U_k^c$ is designed so that it zero forces the signals from \emph{both} interfering transmitters, i.e.,
\begin{align}\label{eq:ZFUc}
\mathbf{U}_{k}^{c}\big[\begin{matrix} \mathbf{H}_{k(k+1)} & \mathbf{H}_{k(k-1)}\end{matrix} \big]=\mathbf{0}.
\end{align}
Since $\mathrm{rank}(\mathbf{H}_{k(k+1)})=D_1$, $\mathrm{rank}(\mathbf H_{k(k-1)})=D_2$, and the channels are independent and randomly generated, with probability $1$, we have
\begin{align}
\mathrm{rank}\big[\begin{matrix} \mathbf{H}_{k(k+1)} & \mathbf{H}_{k(k-1)}\end{matrix} \big]=\min(N, 2M, D_t)=D_t.
\end{align}
Therefore, the left null space of the concatenated $N\times 2M$ matrix $\big[\begin{matrix} \mathbf{H}_{k(k+1)} & \mathbf{H}_{k(k-1)}\end{matrix} \big]$ has dimension $N-D_t$. The rows of $\mathbf{U}_{k}^{c}$ are then selected to be the basis spanning this null space; and hence the size of $\mathbf{U}_{k}^{c}$ is $(N-D_t)\times N$.

The two remaining blocks in $\mathbf R_k$ are designed such that $\mathbf U_{k(k+1)}$ and $\mathbf U_{k(k-1)}$ zero force the interfering signals from transmitter $k+1$ and transmitter $k-1$, respectively, i.e.,
\begin{align}\label{eq:ZFU}
\mathbf{U}_{k(k+1)}\mathbf{H}_{k(k+1)}=\mathbf{0}, \ \mathbf{U}_{k(k-1)}\mathbf{H}_{k(k-1)}=\mathbf{0}.
\end{align}
Furthermore, since $\mathbf U_k^c$ satisfying \eqref{eq:ZFUc} also satisfies the zero-forcing conditions in \eqref{eq:ZFU}, to ensure the invertibility of $\mathbf R_k$, 
the rows of $\mathbf U_{k(k+1)}$ are then selected from the left null space of $\mathbf H_{k(k+1)}$, but \emph{not} in the subspace spanned by the rows of $\mathbf{U}_{k}^{c}$. One possible design is to set $\mathbf U_{k(k+1)}^T$ to be the basis spanning the subspace $\mathcal{N}\big(\mathbf H_{k(k+1)}^T\big) \bigcap \mathcal{N}\big(\mathbf U_k^c\big)$. With $\mathrm{rank} (\mathbf H_{k(k+1)})=D_1$, the dimension of the subspace where $\mathbf U_{k(k+1)}$ can be selected from is then given by $(N-D_1)-(N-D_t)=D_2$. Therefore, the size of $\mathbf U_{k(k+1)}$ is given by $D_2\times N$. $\mathbf U_{k(k-1)}$ can be  designed similarly and its size is $D_1\times N$. Under the assumption that the channel matrices are independent and randomly generated, it follows that with probability 1, the resulting  $N\times N$ matrix $\mathbf R_k$ is of full rank, and hence invertible.

After $\mathbf R_k$ at receiver $k$ is determined, we switch to the design of the invertible linear transformation $\mathbf T_k\in \mathbb{C}^{M\times M}$ at transmitter $k$, which is  partitioned as
\begin{equation}\label{eq:precoder3}
\mathbf{T}_k=\left[\begin{matrix}\mathbf{V}_{(k-1)k} & \mathbf{G}_{k}  & \mathbf{V}_{(k+1)k}\end{matrix}\right],
\end{equation}
where $\mathbf{G}_{k}$ is designed so that it zero forces the interfering signals to \emph{both} receiver $k-1$ and $k+1$, i.e.,
\begin{equation}\label{eq:ZFG}
\left[\begin{matrix}\mathbf{H}_{(k-1)k}\\ \mathbf{H}_{(k+1)k}
\end{matrix} \right]\mathbf G_k=\mathbf 0.
\end{equation}
The columns of $\mathbf G_k$ are then selected to be the basis spanning the null space of the concatenated matrix in \eqref{eq:ZFG}. Since $\mathrm {rank}( \mathbf{H}_{(k-1)k})=D_1$ and $\mathrm {rank}( \mathbf{H}_{(k+1)k})=D_2$,  we have
\[
\mathrm {rank}\left[\begin{matrix}\mathbf{H}_{(k-1)k}\\ \mathbf{H}_{(k+1)k}
\end{matrix} \right]=\min(2N, M, D_t)=D_t.
\]
 Therefore, the null space of the concatenated $2N\times M$ matrix in \eqref{eq:ZFG} has dimension
$M-D_t$ and the size of $\mathbf G_k$ is  $M \times (M-D_t)$.

Next, the remaining two blocks $\mathbf{V}_{(k-1)k}$ and $\mathbf{V}_{(k+1)k}$ in $\mathbf T_k$ are designed so that they zero force the interfering signals to receiver  $k-1$ and receiver $k+1$, respectively, i.e.,
\begin{align}\label{eq:ZFV}
&\mathbf{H}_{(k-1)k}\mathbf{V}_{(k-1)k}=\mathbf{0}, \quad \mathbf{H}_{(k+1)k}\mathbf{V}_{(k+1)k}=\mathbf{0}.
\end{align}
Furthermore, since $\mathbf G_k$ satisfying \eqref{eq:ZFG} also satisfies the zero-forcing conditions in \eqref{eq:ZFV}, to ensure the invertibility of $\mathbf T_k$, 
the columns of $\mathbf{V}_{(k-1)k}$ are chosen from the null space of $\mathbf H_{(k-1)k}$, but \emph{not} in the subspace spanned by the columns of $\mathbf G_k$, e.g., $\mathbf V_{(k-1)k}$ can be set to be the basis spanning the subspace $\mathcal{N}(\mathbf H_{(k-1)k})\bigcap \mathcal{N}(\mathbf G_k^T)$. With $\mathrm{rank}(\mathbf H_{(k-1)k})=D_1$, the dimension of the subspace where  the columns of $\mathbf {V}_{(k-1)k}$ can be selected from is then given by $(M-D_1)-(M-D_t)=D_2$; and hence $\mathbf {V}_{(k-1)k}$  is of size $M\times D_2$. A similar design for $\mathbf {V}_{(k+1)k}$ can be obtained and its size is $M\times D_1$. Under the assumption that the channel matrices are independent and randomly generated, the resulting $M\times M$ matrix $\mathbf T_k$ is of full rank, and hence is invertible.

\begin{figure}[htb]
\centering
\includegraphics[scale=0.65]{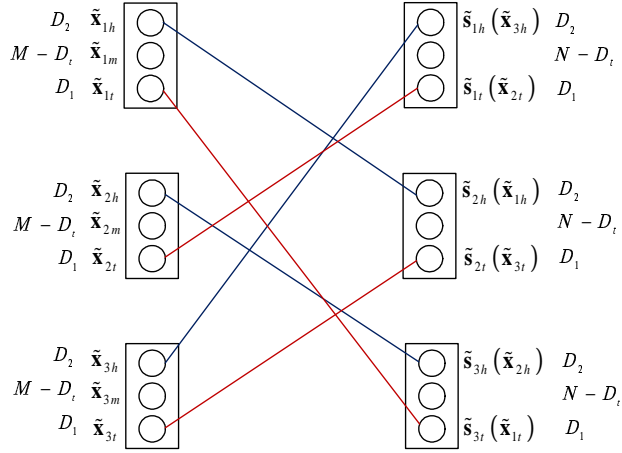}
\caption{The equivalent partially-connected full-rank  MIMO-IC for the low-interference case.\vspace{-5ex}}
\label{F:CoBL}
\end{figure}

%
Next, we derive the resulting channel model with the linear transformations $\mathbf T_k$ and $\mathbf R_k$ applied at the transmitters and receivers, respectively. Denote by $\xtd_k\in \mathbb{C}^{M\times 1}$ the signal vector at transmitter $k$ before applying $\mathbf T_k$. Then the transmitted signal $\mathbf x_k$ is given by $\mathbf x_k=\mathbf T_k \xtd_k$. Based on the partition of $\mathbf T_k$ in \eqref{eq:precoder3}, $\xtd_k$ is decomposed into head, middle and tail parts as $\xtd_k=\left[\begin{matrix}\xtd_{kh}^T& \xtd_{km}^T & \xtd_{kt}^T \end{matrix} \right]^T$, where the dimensions of the three blocks are respectively given by $|\xtd_{kh}|=D_2$, $|\xtd_{km}|=M-D_t$, and $|\xtd_{kt}|=D_1$. Further denote by $\ytd_k\in \mathbb{C}^{N\times 1}$ the signal vector at receiver $k$ after applying linear transformation $\mathbf R_k$, i.e., $\ytd_k =\mathbf R_k \mathbf y_k$, where $\mathbf y_k$ is the received vector at receiver $k$. By substituting $\mathbf y_k$ with \eqref{eq:channelModel} and using $\mathbf x_k=\mathbf T_k \xtd_k$, we get the input-output relationship after the linear transformations $\mathbf T_k$ and $\mathbf R_k$ as
\begin{equation}
\begin{aligned}
\ytd_k=&\mathbf R_k \mathbf H_{kk} \mathbf T_k \xtd_{k}+\mathbf R_k\mathbf n_k+\underbrace{\mathbf R_k \left(\mathbf H_{k(k-1)}\mathbf T_{k-1} \xtd_{k-1}+ \mathbf H_{k(k+1)}\mathbf T_{k+1}\xtd_{k+1}\right)}_{\std_k},
\end{aligned}
\end{equation}
where we have denoted by $\std_k$ the interference vector at receiver $k$. By substituting  $\mathbf R_k$ with \eqref{eq:beamformer3} and $\mathbf T_k$ with \eqref{eq:precoder3}, $\std_k$ can be written as $\std_k=\left[\begin{matrix} \std_{kh}^T & \mathbf 0^T& \std_{kt}^T \end{matrix}\right]^T$, where
\begin{align}
\std_{kh}&= \underbrace{\mathbf U_{k(k+1)}\mathbf H_{k(k-1)}\mathbf{V}_{(k+1)(k-1)}}_{\Htd_{kh}:\ D_2 \times D_2}\xtd_{(k-1)h}\label{eq:skh}
\end{align}
\begin{align}
\std_{kt}&= \underbrace{\mathbf U_{k(k-1)}\mathbf H_{k(k+1)}\mathbf V_{(k-1)(k+1)}}_{\Htd_{kt}: \ D_1 \times D_1}\xtd_{(k+1)t}\label{eq:skt}
\end{align}
The above result shows that by applying the carefully designed linear transformations $\mathbf T_k$ and $\mathbf R_k$, the middle $N-D_t$ antennas of receiver $k$ is interference-free, the head $D_2$ antennas are interfered only by the head $D_2$ antennas from transmitter $k-1$ via matrix $\Htd_{kh}$, and the tail $D_1$ antennas are interfered only by the tail $D_1$ antennas from transmitter $k+1$ via matrix $\Htd_{kt}$. Furthermore, since the channels are independent and randomly generated, with probability $1$, the $D_2 \times D_2$ matrix $\Htd_{kh}$ and the $D_1\times D_1$ matrix $\Htd_{kt}$ are both of full rank and hence invertible. Such an equivalent partially-connected full-rank  MIMO-IC for the low-interference case is illustrated in Fig.~\ref{F:CoBL}.\footnote{Note that only the residue interfering links are shown in Fig.~\ref{F:CoBL} and in all similar plots hereafter. The direct links are omitted for conciseness.}

\subsection{High-Interference Case: $M<D_t\leq 2M$}\label{sec:CoBHigh}
 In the high-interference case where the interfering channel matrices have higher ranks, more residue interfering links than that shown in Fig.~\ref{F:CoBL} are expected since in this case, the left and right nullspace of the interfering channel matrices from which the zero-forcing vectors are selected have smaller dimensions. As a result, $\mathbf R_k$ and $\mathbf T_k$ need to be designed differently from that given in the previous subsection. Firstly, $\mathbf R_k\in \mathbb{C}^{N\times N}$ at receiver $k$ is partitioned as
\begin{align}\label{eq:RkH}
\mathbf{R}_{k}=\left[\begin{matrix}\mathbf{U}_{k(k+1)} \\ \mathbf{J}_{k} \\ \mathbf{U}_{k(k-1)}\end{matrix}\right],
\end{align}
where $\mathbf{U}_{k(k+1)}\in \mathbb{C}^{(M-D_1)\times N}$,  $\mathbf U_{k(k-1)}\in \mathbb{C}^{(M-D_2)\times N}$, and $\mathbf J_k \in \mathbb{C}^{(N+D_t-2M)\times N}$. The rows of $\mathbf U_{k(k+1)}$ and $\mathbf U_{k(k-1)}$ are selected from the left null spaces of $\mathbf H_{k(k+1)}$ and $\mathbf H_{k(k-1)}$, respectively, but \emph{not} in the intersection of these two null spaces. With $D_t>M$, it can be verified that the specified number of rows for $\mathbf U_{k(k+1)}$ and $\mathbf U_{k(k-1)}$ can always be found. Furthermore, the central block $\mathbf{J}_{k}$  is generated randomly. It then follows that the resulting $N\times N$ matrix $\mathbf R_k$ is full rank, and hence invertible.

With $\mathbf R_k$ specified, we switch to the linear transformation $\mathbf T_k\in\mathbb{C}^{M\times M}$ at transmitter $k$, which is partitioned as
\begin{align}\label{eq:TkH}
\mathbf{T}_{k}=\left[\begin{matrix}\mathbf{V}_{(k-1)k} & \mathbf{Q}_{k} & \mathbf{V}_{(k+1)k} \end{matrix}\right],
\end{align}
where $\mathbf V_{(k-1)k}\in \mathbb{C}^{M\times {(M-D_1)}}$ and $\mathbf V_{(k+1)k}\in \mathbb{C}^{M\times {(M-D_2)}}$ are selected to be the basis spanning the null spaces of $\mathbf H_{(k-1)k}$ and $\mathbf H_{(k+1)k}$, respectively. Note that since $D_t>M$, we have $(M-D_1)+(M-D_2)<M$. As a result, these two null spaces have empty intersection, and hence the columns of $\mathbf V_{(k-1)k}$ and $\mathbf V_{(k+1)k}$ are linearly independent. Next, the central block $\mathbf Q_k$ of $\mathbf{T}_k$ is designed so that, together with the receiver matrix $\mathbf R_{k-1}$ or $\mathbf R_{k+1}$, the interferences to a particular subset of antennas of receiver $k-1$ and receiver $k+1$ are zero-forced. More specifically, we need to have
\begin{align}\label{eq:Qk}
\left[\begin{matrix}\mathbf{U}_{(k-1)(k+1)}\mathbf{H}_{(k-1)k} \\
\mathbf{U}_{(k+1)(k-1)}\mathbf{H}_{(k+1)k}
\end{matrix}\right]\mathbf{Q}_{k}=\mathbf{0}.
\end{align}
Since $\mathbf{U}_{(k-1)(k+1)}$ is designed solely based on $\mathbf{H}_{(k-1)(k+1)}$, it is independent of $\mathbf{H}_{(k-1)k}$. Thus, we have
\begin{align*}
\mathrm{rank}(\mathbf{U}_{(k-1)(k+1)}\mathbf{H}_{(k-1)k})&=\min(\mathrm{rank}(\mathbf{U}_{(k-1)(k+1)}),\mathrm{rank}(\mathbf{H}_{(k-1)k}))\\
&=\min(M-D_2,D_1)=M-D_2.
\end{align*}
Similarly, we can obtain that $\mathrm{rank}(\mathbf{U}_{(k+1)(k-1)}\mathbf{H}_{(k+1)k})=M-D_1$. Since $\mathbf{U}_{(k-1)(k+1)}\mathbf{H}_{(k-1)k}$ and $\mathbf{U}_{(k+1)(k-1)}\mathbf{H}_{(k+1)k}$ are independent, we have
\begin{align*}
\mathrm{dim}& \left(\mathcal{N}\left(\left[\begin{matrix}\mathbf{U}_{(k-1)(k+1)}\mathbf{H}_{(k-1)k} \\ \mathbf{U}_{(k+1)(k-1)}\mathbf{H}_{(k+1)k}\end{matrix}\right]\right)\right)\\
=&\mathrm{dim}\left(\mathcal{N}(\mathbf{U}_{(k-1)(k+1)}\mathbf{H}_{(k-1)k})\right)+\mathrm{dim}\left(\mathcal{N}(\mathbf{U}_{(k+1)(k-1)}\mathbf{H}_{(k+1)k})\right)-M
=D_t-M.
\end{align*}
Therefore, the columns of $\mathbf{Q}_{k}$ can be chosen to be the basis spanning the null space of the concatenated matrix in \eqref{eq:Qk} and its size is given by $M\times(D_t-M)$. 
It follows that the resulting $M\times M$ matrices $\mathbf T_k$ is full-rank and hence invertible.

Following similar analysis as for the low-interference case, the interference vector $\std_{k}=\left[\begin{matrix}\std_{kh}^T & \std_{km}^T & \std_{kt}^T\end{matrix}\right]^T$ at receiver $k$ with transformations $\mathbf R_k$ and $\mathbf T_k$ can be obtained, where
\begin{align}
\std_{kh}
=&\underbrace{\mathbf{U}_{k(k+1)}\mathbf{H}_{k(k-1)}\mathbf{V}_{(k+1)(k-1)}}_{\HtdH_{kh}: \ (M-D_1)\times (M-D_1)}\xtd_{(k-1)h}\label{eq:HkhH}
\end{align}
\begin{equation}\label{eq:Pij}
\begin{aligned}
\std_{km}
=&\underbrace{\mathbf{J}_{k}\mathbf{H}_{k(k+1)}\mathbf{Q}_{k+1}}_{\mathbf{P}_{k(k+1)}}\xtd_{(k+1)m}
+\underbrace{\mathbf{J}_{k}\mathbf{H}_{k(k-1)}\mathbf{Q}_{k-1}}_{\mathbf{P}_{k(k-1)}}\xtd_{(k-1)m}\\
&+\mathbf{J}_{k}\mathbf{H}_{k(k+1)}\mathbf{V}_{(k-1)(k+1)}\xtd_{(k+1)t}+\mathbf{J}_{k}\mathbf{H}_{k(k-1)}\mathbf{V}_{(k+1)(k-1)}\xtd_{(k-1)h}
\end{aligned}
\end{equation}
\begin{align}
\std_{kt}
=&\underbrace{\mathbf{U}_{k(k-1)}\mathbf{H}_{k(k+1)}\mathbf{V}_{(k-1)(k+1)}}_{\HtdH_{kt}: \ (M-D_2) \times (M-D_2)}\xtd_{(k+1)t}\label{eq:HktH}
\end{align}

\begin{figure}[htb]
\centering
\includegraphics[scale=0.65]{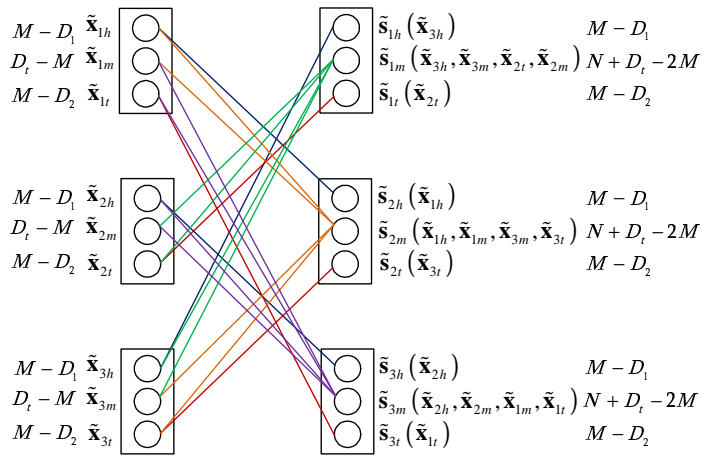}
\caption{The equivalent partially-connected full-rank  MIMO-IC for the high-interference case.\vspace{-3ex}}
\label{F:CoBH}
\end{figure}

It can be obtained that the square matrices $\HtdH_{kh}$ and $\HtdH_{kt}$ defined above are both \emph{full-rank}. Furthermore, the $(N+D_t-2M)\times(D_t-M)$ matrices $\mathbf{P}_{k(k+1)}$ and $\mathbf{P}_{k(k-1)}$ are both of full column rank. The equivalent partially-connected full-rank MIMO-IC with linear transformations $\mathbf R_k$ and $\mathbf T_k$ for the high-interference case is shown in Fig.~\ref{F:CoBH}.


\section{Information Theoretic DoF Outer Bound}\label{sec:outerBound}
In this section, we present the proof for the converse part of Theorem \ref{theo:main}. First, with single-user bounding, we have
\begin{align}\label{eq:dlessD0}
d\leq D_0.
\end{align}

The remaining bounds in Theorem~\ref{theo:main} are obtained by using the genie-aided signaling technique. To identify the appropriate side information to be provided by the genie, we use the equivalent partially-connected full-rank MIMO-IC developed in the preceding section. As pointed out in \cite{362}, since no claim to duality can be made a-priori for information theoretic DoF results, in principle, the two cases $M_T\leq M_R$ and $M_T>M_R$ need to be discussed separately. However, in this paper, we only present the detailed derivations for $M_T\leq M_R$, or $M=M_T$ and $N=M_R$, as the reciprocal channel can be dealt with similarly. Similar as in Section~\ref{sec:changeOfBasis}, we need to distinguish the low- and high-interference cases.

\subsection{Low-Interference Case: $D_t\leq M$}
For the low-interference case, the interfering links of the equivalent partially-connected full-rank  MIMO-IC are plotted in Fig.~\ref{F:CoBL}. Since we are dealing with a converse argument, it follows by assumption that each receiver is able to decode and subtract out its desired signal. Consider the interference vector $\std_{1h}$  received by the head antennas of receiver $1$. As given in \eqref{eq:skh}, $\std_{1h}$ is a function of $\xtd_{3h}$ via the $D_2\times D_2$ matrix $\Htd_{1h}$. Since $\Htd_{1h}$ is full rank with probability $1$, receiver $1$ is able to distinguish $\xtd_{3h}$ based on $\std_{1h}$ subject to the noise distortion. 
 With a similar argument, receiver $1$ is also able to demodulate $\xtd_{2t}$ subject to the noise distortion after subtracting its desired signal. Hence, if a genie provides the signal $G_{1L}=\{\xtd_{2h}^{n}, \xtd_{2m}^{n},\xtd_{3m}^{n}, \xtd_{3t}^{n}\}$ to receiver $1$, receiver $1$ should be able to decode all the three messages $W_1$, $W_2$ and $W_3$ subject to the noise
distortion. Let $R_k$ denote the rate for user $k$ and $R_{\Sigma}$ the sum rate of the three users. By Fano's inequality, we have
\begin{align}
nR_{\Sigma}\leq &\ I(W_1, W_2, W_3; \ytd_{1}^n, G_{1L})+ \smallterm \notag \\ 
 \leq &h(\ytd_{1}^{n},G_{1L})- h(\ytd_{1}^{n},G_{1L}|W_1, W_2, W_3)+\smallterm \notag \\
 \leq & h(\ytd_{1}^{n})+h(G_{1L}|\ytd_{1}^{n})+\smallterm \nonumber\\
\overset{(a)}{\leq}
&Nn\log\rho+h(\xtd_{2h}^{n}, \xtd_{2m}^{n}, \xtd_{3m}^{n},\xtd_{3t}^{n}|\ytd_{1}^{n}) +\smallterm \nonumber \\
\leq
&Nn\log\rho+h(\xtd_{2h}^{n}, \xtd_{2m}^{n}|\ytd_{1}^{n})+h(\xtd_{3m}^{n},\xtd_{3t}^{n}|\ytd_{1}^{n})+\smallterm \nonumber\\
\overset{(b)}{\leq}
&Nn\log\rho+h(\xtd_{2h}^{n}, \xtd_{2m}^{n}|\xtd_{2t}^{n})+h(\xtd_{3m}^{n},\xtd_{3t}^{n})+\smallterm
\nonumber\\
\overset{(c)}{\leq}
&Nn\log\rho+nR_2-h(\xtd_{2t}^{n})+h(\xtd_{3m}^{n},\xtd_{3t}^{n})+\smallterm\label{eq:OuterL1},
\end{align}
where $(a)$ is true since receiver $1$ has only $N$ antennas, $(b)$ follows since conditioning does not increase the entropy, and $(c)$ is true as the message $W_2$ can be decoded from the observation of $\{\xtd_{2h}^n,\xtd_{2m}^n,\xtd_{2t}^n\}$.

By symmetry, all three messages can be decoded subject to the noise distortion if a genie provides the signals $G_{2L}=\{\xtd_{3h}^{n}, \xtd_{3m}^{n}, \xtd_{1m}^{n},\xtd_{1t}^{n}\}$ and $G_{3L}=\{\xtd_{1h}^{n}, \xtd_{1m}^{n}, \xtd_{2m}^{n},\xtd_{2t}^{n}\}$ to receiver $2$ and receiver $3$, respectively. We thus have
\begin{align}
nR_{\Sigma}\leq& Nn\log\rho+nR_3-h(\xtd_{3t}^{n})+h(\xtd_{1m}^{n},\xtd_{1t}^{n}) +\smallterm \label{eq:OuterLR2}\\
nR_{\Sigma}\leq& Nn\log\rho+nR_1-h(\xtd_{1t}^{n})+h(\xtd_{2m}^{n},\xtd_{2t}^{n})+\smallterm \label{eq:OuterLR3}
\end{align}

By summing up \eqref{eq:OuterL1}-\eqref{eq:OuterLR3}, we get
\begin{align}
\small
2nR_{\Sigma}\leq& 3Nn\log\rho+\sum_{k=1}^{3}h(\xtd_{km}^{n}|\xtd_{kt}^{n})+\smallterm \nonumber\\
\leq&3Nn\log\rho+\sum_{k=1}^{3}h(\xtd_{km}^{n})+\smallterm\nonumber\\
\overset{(d)}{\leq}&3Nn\log\rho+3(M-D_t)n\log\rho+\smallterm,\label{eq:OuterLRoverall}
\end{align}
where $(d)$ follows since the dimension of $\xtd_{km}$ is  $M-D_t$ as given in Section~\ref{sec:CoBLow}.

Let $R$ denote the rate normalized per user, i.e., $R=R_{\Sigma}/3$. Then \eqref{eq:OuterLRoverall} can be equivalently written as
\begin{align}\label{eq:OuterLRoverall2}
6nR\leq & 3N n \log \rho +3(M-D_t)n\log \rho+\smallterm.
\end{align}
By dividing $\log\rho$ and $n$ on both sides of \eqref{eq:OuterLRoverall2}, and letting $\rho\rightarrow\infty$ and $n\rightarrow\infty$, we obtain the following DoF outer bound for the low-interference case:
\begin{align}
d\leq\frac{N+M-D_t}{2}. \label{eq:DoFOuterL}
\end{align}

\subsection{High-Interference Case: $M<D_t\leq 2M$}
For the high-interference case, the interfering links of the equivalent channel after invertible linear transformations are plotted in Fig.~\ref{F:CoBH}. Consider the interference vector $\std_{1h}$ received by the head antennas of receiver $1$, which is a function of $\xtd_{3h}$ via the square matrix $\HtdH_{1h}$ as shown in \eqref{eq:HkhH}. Since $\HtdH_{1h}$ is full rank,  after decoding and subtracting  its desired signal, receiver $1$ is able to distinguish $\xtd_{3h}$ subject to the noise distortion based on the observation of $\std_{1h}$. Similarly, receiver $1$ can also demodulate $\xtd_{2t}$ subject to the noise distortion. With  $\xtd_{3h}$ and $\xtd_{2t}$ obtained, the interference terms at the middle antennas of receiver $1$ due to these two signals can be subtracted out from $\std_{1m}$,  leaving it a linear function of  $\xtd_{2m}$ and $\xtd_{3m}$ via the full-column rank matrices $\mathbf{P}_{12}$ and $\mathbf{P}_{13}$, respectively.  Therefore, if receiver $1$ is provided with the side information $\xtd_{2m}$, it will be able to get $\xtd_{3m}$ subject to the noise distortion; and vice versa.

Based on the above analysis, if a genie  provides the signals $G_{1H}=\{\xtd_{2h}^{n},\xtd_{2m}^{n},\xtd_{3t}^{n}\}$ to receiver $1$, receiver $1$ would be able to decode all the three messages subject to the noise
distortion. With  Fano's inequality, we have
\begin{align}
\hspace{-1ex} nR_{\Sigma}\leq & I(W_1, W_2, W_3; \ytd_1^n, G_{1H})+\smallterm\notag \\
\leq & h(\ytd_1^n) +h(G_{1H}|\ytd_1^n)+\smallterm\notag \\
\leq &
Nn\log\rho+h(\xtd_{2h}^{n},\xtd_{2m}^{n},\xtd_{3t}^{n}|\ytd_{1}^{n})+\smallterm\notag\\
=&
Nn\log\rho+h(\xtd_{2h}^{n},\xtd_{2m}^{n}|\xtd_{3t}^{n},\ytd_{1}^{n})
+h(\xtd_{3t}^{n}|\ytd_{1}^{n})+\smallterm\notag \\
\leq &
Nn\log\rho+h(\xtd_{2h}^{n},\xtd_{2m}^{n}|\xtd_{2t}^{n})
+h(\xtd_{3t}^{n})  +\smallterm\notag \\
\leq &
Nn\log\rho+nR_2-h(\xtd_{2t}^{n})+h(\xtd_{3t}^{n})+\smallterm\label{eq:OuterH1f}
\end{align}

Due to symmetry, by advancing the user indices to receiver $2$ and receiver $3$, the following inequalities can be obtained:
\begin{align}
nR_{\Sigma}
\leq &
Nn\log\rho+nR_3-h(\xtd_{3t}^{n})+h(\xtd_{1t}^{n})+\smallterm\label{eq:OuterH2f}\\
nR_{\Sigma}
\leq &
Nn\log\rho+nR_1-h(\xtd_{1t}^{n})+h(\xtd_{2t}^{n})+\smallterm\label{eq:OuterH3f}
\end{align}
By summing up \eqref{eq:OuterH1f}-\eqref{eq:OuterH3f} and  with $R_{\Sigma}=3R$, we have
\begin{align}
6nR\leq
&3Nn\log\rho+\smallterm\label{eq:OuterHoverall1}
\end{align}
By dividing $\log\rho$ and $n$ on both sides of \eqref{eq:OuterHoverall1}, and letting $\rho\rightarrow\infty$ and $n\rightarrow\infty$, we obtain the following DoF outer bound for the high-interference case:
\begin{align}
d\leq\frac{N}{2}\label{eq:DoFOuterH1}.
\end{align}

To prove the remaining two bounds in Theorem~\ref{theo:main}, i.e.,
\begin{align}\label{eq:pbound}
d\leq\min\left(\frac{pM}{2p-1},\frac{pN+2M-D_t}{2p+1}\right),
\end{align}
where $p=\left\lceil\frac{D_t-M}{N-M}\right\rceil$, 
we will first give the detailed derivations for $p=1$ and $p=2$, followed by a brief description for general $p$ values as the main technique follows from  \cite{362}.

\subsubsection{$p=1$}
For $p=\big\lceil\frac{D_t-M}{N-M}\big\rceil=1$, we have $0<\frac{D_t-M}{N-M}\leq 1$, or equivalently
\begin{align}\label{eq:p1condition}
M<D_t\leq N.
\end{align}
In this case, we need to show that $d\leq\min\left(M,\frac{N+2M-D_t}{3}\right)$ according to \eqref{eq:pbound}. Note that the bound $d\leq M$ follows trivially from the single-user bound shown in \eqref{eq:dlessD0} since $D_0\leq M$. To show  $d\leq \frac{N+2M-D_t}{3}$ under the condition given in \eqref{eq:p1condition},  we again use the equivalent channel given in Fig.~\ref{F:CoBH}.

As discussed previously, after subtracting its desired signal, receiver $1$ is able to obtain $\xtd_{3h}$ and $\xtd_{2t}$ subject to the noise distortion. Therefore, the interference caused  by these two signals can be subtracted out from $\std_{1m}$, leaving the middle antennas interfered by $\xtd_{2m}$ and $\xtd_{3m}$ only via full-column rank matrices. Furthermore, we have $|\xtd_{2m}|+|\xtd_{3m}|=2(D_t-M)$, which is no greater than $|\std_{1m}|=N+D_t-2M$ under the condition given in \eqref{eq:p1condition}. Therefore, $\xtd_{2m}$ and $\xtd_{3m}$ are also  distinguishable by receiver $1$ subject to the noise distortion. 
As a result, if a genie provides $\{\xtd_{2h}^{n},\xtd_{3t}^{n}\}$ to receiver $1$, receiver $1$ will be able to decode all the three messages subject to the noise
distortion. Therefore, we have
\begin{align}
3nR&\leq I(W_1, W_2, W_3; \ytd_1^n,\xtd_{2h}^{n},\xtd_{3t}^{n}) +\smallterm \notag\\
&\leq h(\ytd_1^n) +h(\xtd_{2h}^{n},\xtd_{3t}^{n}|\ytd_1^n) +\smallterm \notag\\
&\leq
Nn\log\rho+h(\xtd_{2h}^{n})+h(\xtd_{3t}^{n})+\smallterm \notag\\
&\leq
n(N+2M-D_t)\log\rho+\smallterm,\label{eq:p1N}
\end{align}
where the last inequality follows since $|\xtd_{2h}|=M-D_1$ and $|\xtd_{3t}|=M-D_2$.
\eqref{eq:p1N} leads to the desired DoF outer bound
\begin{align*}
d\leq\frac{N+2M-D_t}{3}.
\end{align*}

To prove the upper bound \eqref{eq:pbound} for $p\geq 2$, we need to further decompose the middle antennas in Fig.~\ref{F:CoBH} (i.e., $\xtd_{km}$ and $\std_{km}$) by making use of the equivalent channel structure developed in \cite{362} for the full-rank  MIMO-IC. Specifically, the middle antennas of each user (antennas $km$ for user $k$ as shown in Fig.~\ref{F:CoBH}), which impose no interference to the head and tail antennas (antennas $kh$ and $kt$ for user $k$), can be first treated separately to form a $3$-user fully-connected \emph{full-rank} MIMO-IC with $\tilde{M}$ transmit antennas and $\tilde N$ receive antennas, where $\tilde M=D_t-M$ and $\tilde N=N+D_t-2M$. As such, the equivalent channel model given in \cite{362} can be applied to the middle antennas, after which the outer layer formed by the head and tail antennas ($kh$ and $kt$ for user $k$) can be integrated. We start with the detailed discussion for $p=2$ in the following.

\subsubsection{$p=2$}
\begin{figure}[!htb]
\centering
\includegraphics[scale=1.1]{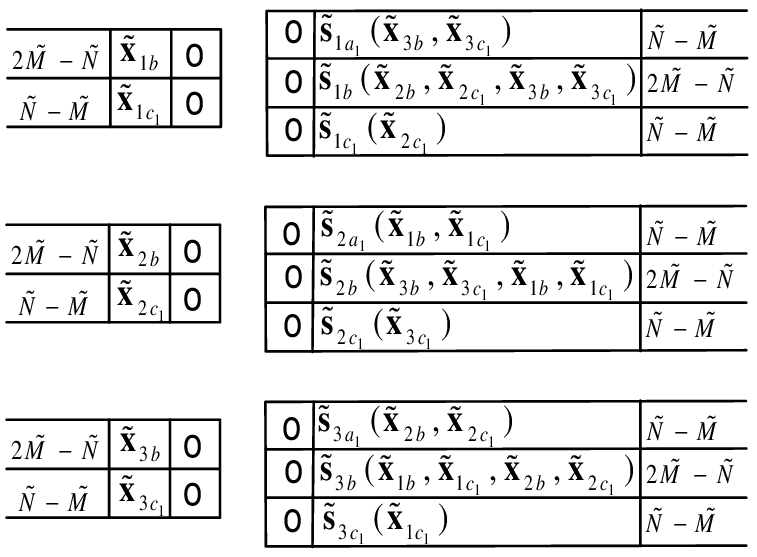}
\caption{An equivalent channel for the $3$-user $\tilde{M} \times \tilde N$ full-rank MIMO-IC with $\tilde M/\tilde N \in (1/2, 2/3]$.\vspace{-2ex}}
\label{F:P2FullRank}
\end{figure}

\begin{figure}[!htb]
\centering
\includegraphics[scale=0.6]{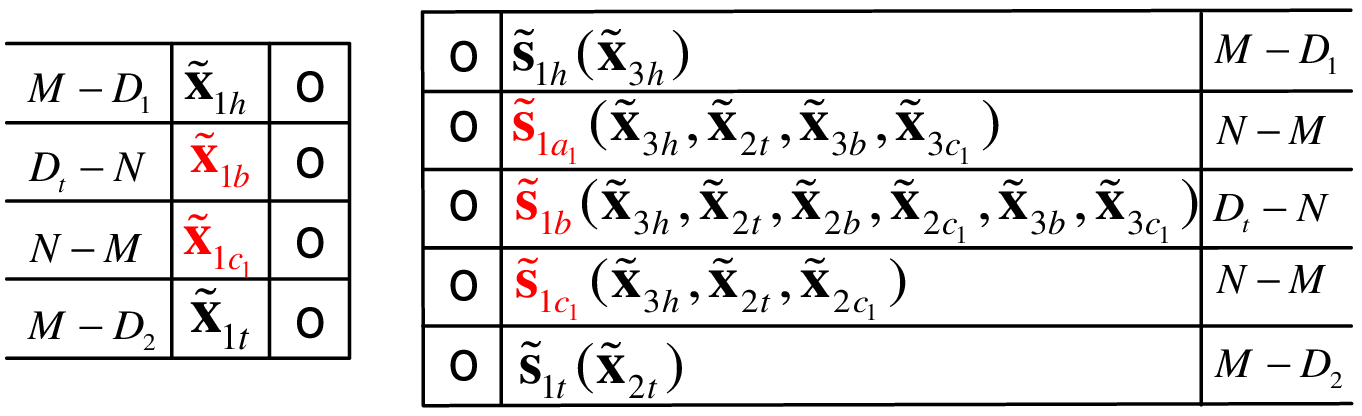}
\caption{An equivalent channel for the $3$-user rank-deficient MIMO-IC when $p=2$.\vspace{-4ex}}
\label{F:Pequals2}
\end{figure}

For $p=\big\lceil\frac{D_t-M}{N-M}\big\rceil=\big\lceil\frac{\tilde M}{\tilde N- \tilde M}\big\rceil=2$, we have
\begin{align}\label{eq:p2condition}
N<D_t\leq 2N-M,
\end{align}
or $\tilde M/\tilde N \in (1/2, 2/3]$. For $3$-user $\tilde{M} \times \tilde N$ full-rank MIMO-IC with $\tilde M/\tilde N \in (1/2, 2/3]$, an equivalent channel obtained by appropriate linear transformations 
is given in Appendix~A.2.2 of \cite{362}, which is reproduced in Fig.~\ref{F:P2FullRank}.\footnote{Note that a slightly different form from that in \cite{362} is given in Fig.~\ref{F:P2FullRank}, where the transmitted signal by user $k$ is decomposed into $\xtd_{kb}$ and $\xtd_{kc_1}$ for easier reference.} It is observed that in the equivalent channel, the bottom $\tilde N-\tilde M$ antennas of receiver $k$ (i.e., $\std_{kc_1}$) are interfered only by the last $\tilde N-\tilde M$ antennas of transmitter $k+1$ (i.e. $\xtd_{(k+1)c_1}$); and the top $\tilde N-\tilde M$ antennas of receiver $k$ (i.e., $\std_{ka_1}$) are interfered by transmitter $k-1$ only. By applying this equivalent channel to the $\tilde M\times \tilde N$ full-rank MIMO-IC formed by the middle antennas of Fig.~\ref{F:CoBH}, and after combining with the head and tail antennas, we obtain an equivalent channel for the rank-deficient MIMO-IC corresponding to $p=2$, as shown in Fig.~\ref{F:Pequals2}.\footnote{Note that due to symmetry, we only show the signals for user $1$ in Fig.~\ref{F:Pequals2} and all similar figures hereafter.} 

 The DoF outer bound in \eqref{eq:pbound} corresponding to $p=2$ is then derived based on Fig.~\ref{F:Pequals2}.  Firstly, $\xtd_{3h}$ and $\xtd_{2t}$ can be demodulated by receiver $1$ subject to the noise distortion, and hence can be subtracted out from $\std_{1c_1}$. This in turn makes $\xtd_{2c_1}$ distinguishable as well. Furthermore, if a genie provides $\xtd_{2b}$ to receiver 1, the two remaining signals $\xtd_{3b}$ and $\xtd_{3c_1}$ can be demodulated based on $\std_{1a_1}$ and $\std_{1b}$. As a result, receiver $1$ will be able to decode all the three messages subject to the noise
distortion if it is provided with the side information $\{\xtd_{2b}^n,\xtd_{2h}^n,\xtd_{3t}^n\}$. With Fano's inequality, we have
\begin{align}
3nR
\leq& I(W_1, W_2, W_3; \ytd_1^n,\xtd_{2b}^{n},\xtd_{2h}^{n},\xtd_{3t}^{n})+\smallterm \notag\\
\leq& h(\ytd_1^n) +h(\xtd_{2b}^{n},\xtd_{2h}^{n},\xtd_{3t}^{n}|\ytd_1^n) +\smallterm\notag\\
\leq&
Nn\log\rho+h(\xtd_{3t}^{n}|\ytd_1^n,\xtd_{2b}^{n},\xtd_{2h}^{n})+h(\xtd_{2b}^{n},\xtd_{2h}^{n}|\ytd_1^n)+\smallterm\notag\\
\leq&
Nn\log\rho+h(\xtd_{3t}^{n}|\xtd_{3h}^{n},\xtd_{3b}^{n},\xtd_{3c_1}^{n})+h(\xtd_{2b}^{n},\xtd_{2h}^{n}|\xtd_{2c_1}^{n}) +\smallterm \notag\\
\leq&
Nn\log\rho+nR_3-h(\xtd_{3h}^{n},\xtd_{3b}^{n},\xtd_{3c_1}^{n})+h(\xtd_{2b}^{n},\xtd_{2h}^{n}|\xtd_{2c_1}^{n})+\smallterm.\label{eq:p2Mu1}
\end{align}

By advancing user indices, we can obtain two other inequalities similar to \eqref{eq:p2Mu1} at receiver 2 and 3, respectively. By summing them up, we get
\begin{align}
6nR\leq& 3Nn\log\rho-\sum_{k=1}^{3}h(\xtd_{kc_1}^{n})+\smallterm.\label{eq:p2Muk}
\end{align}
Moreover, we have
\begin{align}
3nR\leq &
\sum_{k=1}^{3}h(\xtd_{kh}^{n},\xtd_{kb}^{n},\xtd_{kc_1}^{n},\xtd_{kt}^{n})+\smallterm\notag\\
\leq& \sum_{k=1}^{3}\left[h(\xtd_{kh}^{n},\xtd_{kb}^{n},\xtd_{kt}^{n})+h(\xtd_{kc_1}^{n})\right]+\smallterm\notag\\
\leq&3(2M-N)n\log\rho+\sum_{k=1}^{3}h(\xtd_{kc_1}^{n})+\smallterm.\label{eq:p2Muk2}
\end{align}
By summing up \eqref{eq:p2Muk} and \eqref{eq:p2Muk2}, we have
\begin{equation}\label{eq:p2Mfinal}
9nR\leq 6Mn\log\rho+\smallterm,
\end{equation}
which gives the following DoF outer bound:
\begin{equation}\label{E:p2bound1}
d\leq \frac{2M}{3}.
\end{equation}


Furthermore, it can be verified that under  condition  \eqref{eq:p2condition}, receiver $1$ will be able to decode all three messages subject to the noise distortion if a genie provides the signal $\{\xtd_{3c_1}^n,\xtd_{2h}^n,\xtd_{3t}^n\}$. Then with Fano's inequality, we have
\begin{align}
3nR
\leq& I(W_1, W_2, W_3; \ytd_1^n,\xtd_{3c_1}^{n},\xtd_{2h}^{n},\xtd_{3t}^{n})+\smallterm \notag\\
\leq &
h(\ytd_1^n) +h(\xtd_{3c_1}^{n},\xtd_{2h}^{n},\xtd_{3t}^{n}|\ytd_1^n)+\smallterm \notag\\
\leq &
Nn\log\rho+h(\xtd_{2h}^{n}|\ytd_1^n,\xtd_{3c_1}^{n},\xtd_{3t}^{n})+h(\xtd_{3c_1}^{n},\xtd_{3t}^{n}|\ytd_1^n) +\smallterm \notag\\
\leq &
Nn\log\rho+h(\xtd_{2h}^{n}|\xtd_{2c_1}^{n},\xtd_{2t}^{n})+h(\xtd_{3c_1}^{n},\xtd_{3t}^{n})+\smallterm.\label{eq:p2Nu1}
\end{align}
By advancing user indices, we can obtain two other inequalities similar to \eqref{eq:p2Nu1}. By summing them up, we get
\begin{align}\label{eq:p2Neq1}
9nR\leq & 3Nn\log\rho+\sum_{k=1}^{3}h(\xtd_{kh}^{n},\xtd_{kc_1}^{n},\xtd_{kt}^{n})+\smallterm.
\end{align}

Summing up \eqref{eq:p2Muk} and \eqref{eq:p2Neq1} yields
\begin{align}
\hspace{-1ex} 15nR\leq & 6Nn\log\rho+\sum_{k=1}^{3}h(\xtd_{kh}^{n},\xtd_{kt}^{n}|\xtd_{kc_1}^{n})+\smallterm \notag\\
\leq &
6Nn\log\rho+3(2M-D_t)+\smallterm,
\end{align}
which gives the following DoF outer bound:
\begin{align}\label{E:p2bound2}
d\leq\frac{2N+2M-D_t}{5}.
\end{align}
By combining \eqref{E:p2bound1} and \eqref{E:p2bound2}, the upper bound \eqref{eq:pbound} for $p=2$ follows.

\subsubsection{Intuition of ``onion peeling'' for general $p$ values}
We have shown the bound \eqref{eq:pbound} for $p=1,2$. The proof for general $p$ values can be obtained similarly. Since the mathematical derivations overlap largely with the full-rank case \cite{362}, we only give a brief description of the techniques used. It is not difficult to see that for all the $p$ values discussed, the proof of \eqref{eq:pbound} follows the ``onion peeling'' intuition, as termed in \cite{362}. Specifically, depending on the relationship between $D_t$, $M$ and $N$, an equivalent channel with \emph{layered} structure is developed, where the antennas in a particular layer are only interfered by the outer layers, but not by the inner layers. For instance, the signals shown in red in Fig.~\ref{F:Pequals2} can be viewed as the innermost layer and those in black as the outermost layer. With the equivalent layered channel structure, the genie information is provided such that the signals in the outer layer can be always demodulated prior to the processing of the received signals in the inner layer. As such, the interference caused by the outer-layer signals can be subtracted out from the inner-layer antennas. This in effect peels out the outer layer antennas; and hence is given the term ``onion peeling''. As $p$ increases, the antenna blocks involving $D_t$ can be further decomposed. Therefore, the number of layer increases  and the onion peeling technique applies recursively. Note that as compared to the full-rank case, the equivalent layered channel structure for the rank-deficient MIMO-IC has one additional layer, i.e., the outermost layer formed by the head and tail antennas, which is obtained by extracting the rank-deficiency of the interfering channel matrices as discussed previously. When reducing to the full-rank case, i.e., $D_1=D_2=M$, the outer layer vanishes as evidenced from Fig.~\ref{F:CoBH}, in which case exactly the same equivalent layered channel given in \cite{362} will be resulted.

\section{Achievability Proof}\label{sec:innerBound}
In this section, we show the achievability of the spatially-normalized DoF given in Theorem~\ref{theo:main}. 
   Due to space limitation and to convey the most essential ideas without involving overcomplicated mathematics, we only show the achievability through linear dimension counting approach. The more rigorous information-theoretic proof can be obtained following similar arguments as \cite{362}, together with the precoding/decoding techniques presented in this section. With the randomness and independence of the channel matrices and the fact that the linear precoding and decoding matrices are designed solely based on the interfering channel matrices as will be given later,  the subspaces occupied by the desired data symbols and that by the interference will have no overlapping if
\begin{align}\label{eq:lessN}
\dofbar +Z\leq N,
\end{align}
where $\dofbar$ is the number of desired data symbols sent by the transmitter, $Z$ is the dimension of the subspace occupied by the interfering signals, and $N$ is the total dimension available at the receiver. If the above relationship holds, then the receiver can discard the $Z$ dimensions that contain interference and the remaining dimensions are enough to decode the $\dofbar$ desired data symbols \cite{365}.

\begin{figure*}
\centering
\includegraphics[scale=0.6]{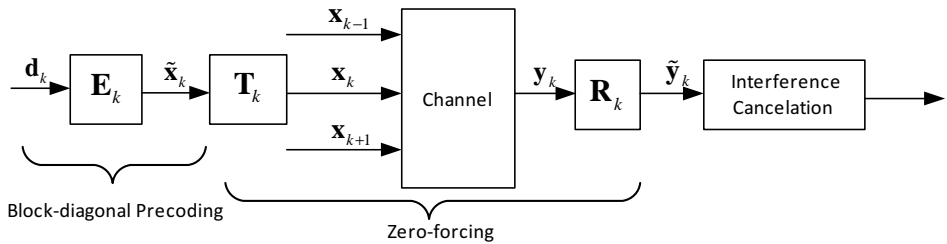}
\caption{The proposed two-layered achievability scheme.\vspace{-4ex}}
\label{F:precoding}
\end{figure*}

 Fig.~\ref{F:precoding} gives an overview of our proposed achievability scheme, which has  a two-layered linear processing structure. The zero forcing achieved by linear transformations $\mathbf R_k$ and $\mathbf T_k$ discussed in Section~\ref{sec:changeOfBasis} constitutes the inner layer, with which  an equivalent partially-connected  full-rank MIMO-IC is obtained, as shown in Fig.~\ref{F:CoBL} and Fig.~\ref{F:CoBH} for the low- and high-interference cases, respectively. The outer layer of the proposed scheme consists of a \emph{block-diagonal} precoding matrix $\mathbf E_k$ at the transmitter and interference cancelation, if necessary, at the receiver. 
For the high-interference case, $\mathbf E_k$ is carefully designed so that interference alignment is achieved for the most general cases; whereas for the low-interference case, a random $\mathbf E_k$ is sufficient since most interfering links are zero forced by the inner layer. In the remaining part of this section, we present the detailed design for the outer precoding matrix $\mathbf E_k$ for both the low- and high-interference cases.  
It is worth mentioning that we assume the DoF quantities in the following are all integer values, so that each signaling block is assigned with integer number of symbols. If this is not the case, a spatial extension by an appropriate factor is required, and all the arguments presented in the following still hold in the spatially-extended channel. After DoF normalization by the same factor, the achievability result for the spatially-normalized DoF in Theorem~\ref{theo:main} follows.

\subsection{Low-Interference Case: $0\leq D_t\leq M$}
For the low-interference case, the spatially-normalized DoF value is given by
 \begin{align}\label{E:termsLow}
 \dofbar=\min\Big(D_0,\frac{N+M-D_t}{2}\Big).
 \end{align}
In the following, we will separately consider the two cases when each term in \eqref{E:termsLow} is active.
 \subsubsection{$\dofbar=\frac{N+M-D_t}{2}$}
 First, we assume that the rank of the direct channel $D_0$ is sufficiently large so that $\dofbar=\frac{N+M-D_t}{2}$. In this case, we have  $\frac{N+M-D_t}{2}\leq D_0\leq M$, which implies $D_t\geq N-M$. We show that each user is able to transmit $\frac{N+M-D_t}{2}$ data symbols, denoted as $\mathbf d_k\in \mathbb{C}^{\dofbar\times 1}$, $k=1,2,3$.

The information-bearing signal vector $\mathbf d_k$ is decomposed as $\mathbf{d}_k=\left[\begin{matrix}\mathbf{d}_{kh}^T \ \mathbf{d}_{km}^T \ \mathbf{d}_{kt}^T\end{matrix}\right]^T$, and it is precoded with a block-diagonal matrix $\mathbf E_k\in\mathbb{C}^{M\times \dofbar}$, which gives
\begin{equation}\label{eq:signalingL}
\small
\xtd_{k}=\left[\begin{matrix}\xtd_{kh}\\ \xtd_{km} \\ \xtd_{kt} \end{matrix} \right]=\left[\begin{matrix}
\mathbf{E}_{kh} & \mathbf{0} & \mathbf{0}\\
\mathbf{0} & \mathbf{E}_{km} & \mathbf{0}\\
\mathbf{0} & \mathbf{0} & \mathbf{E}_{kt}
\end{matrix}\right]
\left[\begin{matrix}\mathbf{d}_{kh}\\ \mathbf{d}_{km}\\ \mathbf{d}_{kt}\end{matrix}\right]
=\left[\begin{matrix}\mathbf{E}_{kh}\mathbf{d}_{kh}\\ \mathbf{E}_{km}\mathbf{d}_{km}\\ \mathbf{E}_{kt}\mathbf{d}_{kt}\end{matrix}\right].
\end{equation}
Based on the partition of $\xtd_k$ as shown in Fig.~\ref{F:CoBL}, the sizes of $\mathbf{E}_{kh}$, $\mathbf{E}_{km}$, $\mathbf{E}_{kt}$  are given by  $D_2\times|\mathbf{d}_{kh}|$, $(M-D_t)\times|\mathbf{d}_{km}|$, and $D_1\times|\mathbf{d}_{kt}|$, respectively, where $|\mathbf{d}_{kh}|, |\mathbf{d}_{km}|$, and $|\mathbf{d}_{kt}|$ are chosen to satisfy
\refstepcounter{equation}\label{eq:condition}
\begin{align}
&|\mathbf{d}_{kh}|+|\mathbf{d}_{km}|+|\mathbf{d}_{kt}|=\dofbar=\frac{N+M-D_t}{2},\tag{\arabic{equation}a} \\
&|\mathbf{d}_{kh}|\leq D_2, \tag{\arabic{equation}b}\\
&|\mathbf{d}_{km}|\leq M-D_t, \tag{\arabic{equation}c}\\
&|\mathbf{d}_{kt}|\leq D_1,  \ k=1,2,3.\tag{\arabic{equation}d}
\end{align}
With each diagonal block in $\mathbf E_k$  randomly generated, the conditions given in \eqref{eq:condition} ensure that $\mathbf E_k$ is of full column rank, which is necessary for the full decodability of the desired symbols. It can be verified that with $D_t\geq N-M$, one feasible choice to satisfy \eqref{eq:condition} is
\begin{equation}\label{eq:dataL}
\begin{aligned}
&|\mathbf{d}_{kh}|=\min\left(D_2,\frac{N-M+D_t}{2}\right),\\
&|\mathbf{d}_{km}|=M-D_t,\\
&|\mathbf{d}_{kt}|=\frac{N-M+D_t}{2}-|\mathbf{d}_{kh}|,  \ k=1,2,3.
\end{aligned}
\end{equation}

Next, we show that the full decodability condition \eqref{eq:lessN} is satisfied. Due to symmetry, we only need to consider receiver $1$.
With zero forcing achieved by $\mathbf T_1$ and $\mathbf R_1$ in the inner layer, it is observed from Fig.~\ref{F:CoBL} that receiver $1$ only sees the interfering signals $\xtd_{3h}$ and $\xtd_{2t}$. Therefore, we have
\begin{align}
Z&=|\xtd_{3h}|+|\xtd_{2t}|=|\mathbf{d}_{3h}|+|\mathbf{d}_{2t}| \notag \\
&=\frac{N-M+D_t}{2}.
\end{align}
As a result, condition \eqref{eq:lessN} is satisfied with equality; and hence $\dofbar=\frac{N+M-D_t}{2}$ is achievable.

 \subsubsection{$\dofbar=D_0$}
In this case, we have $D_0< \frac{N+M-D_t}{2}$. Then the DoF value $D_0$ per user can be achieved by reducing the number of data streams of any information sub-blocks given in \eqref{eq:dataL}, till we have $|\mathbf{d}_{kh}|+|\mathbf{d}_{km}|+|\mathbf{d}_{kt}|=D_0$.

\subsection{High-Interference Case: $M<D_t\leq 2M$}
Similar to the low-interference case, the information-bearing signal  $\mathbf d_k$ for the high-interference case is decomposed as $\mathbf{d}_{k}=\left[\mathbf{d}_{kh}^{T},\mathbf{d}_{km}^{T},\mathbf{d}_{kt}^{T}\right]^T$ and it is precoded with a block-diagonal matrix $\mathbf E_k$ as shown in \eqref{eq:signalingL}. The sizes of $\mathbf E_{kh},\ \mathbf E_{km}$, and $\mathbf E_{kt}$ are now given by
$(M-D_1)\times|\mathbf{d}_{kh}|$, $(D_t-M)\times|\mathbf{d}_{km}|$, and $(M-D_2)\times|\mathbf{d}_{kt}|$, respectively, which correspond to the partition of $\xtd_k$ as in Fig.~\ref{F:CoBH}. To ensure the decodability of $\dofbar$ desired data symbols at each  receiver, the number of  symbols $|\mathbf{d}_{kh}|,|\mathbf{d}_{km}|$, and $|\mathbf{d}_{kt}|$ of each sub-block in $\mathbf d_k$ are chosen to satisfy
\refstepcounter{equation}\label{eq:conditionH}
\begin{align}
&|\mathbf{d}_{kh}|+|\mathbf{d}_{km}|+|\mathbf{d}_{kt}|=\dofbar, \tag{\arabic{equation}a} \\
&|\mathbf{d}_{kh}|\leq M-D_1, \tag{\arabic{equation}b}\\
&|\mathbf{d}_{km}|\leq D_t-M, \tag{\arabic{equation}c}\\
&|\mathbf{d}_{kt}|\leq M-D_2,  \ k=1,2,3.\tag{\arabic{equation}d}
\end{align}
The blocks $\mathbf E_{kh}$ and $\mathbf E_{kt}$ for the head and tail antennas are  randomly generated, so that under the conditions given in \eqref{eq:conditionH}, they are of full column rank almost surely. 
In the remaining part of this subsection, for different $\dofbar$ values activated in \eqref{eq:exactDoF}, we present a feasible selection of $|\mathbf{d}_{kh}|,|\mathbf{d}_{km}|$, and $|\mathbf{d}_{kt}|$ to satisfy \eqref{eq:conditionH}, as well as the designs of the precoding matrices $\mathbf E_{km}$ for the middle antennas so that the full decodability condition given in \eqref{eq:lessN} is satisfied.

Similar to the low-interference case, if the DoF is limited by the direct channel, i.e. $\dofbar=D_0$, the achievability scheme can be simply obtained by reducing the number of data symbols transmitted for the interfering-link-limited cases. Therefore, in the following, we assume without loss of generality that the rank of the direct channel $D_0$ is large enough, i.e., $\frac{N}{2}<D_0\leq M$, so that the achievable DoF $\dofbar$ is limited by the interfering links. 
  In this case,  it can be verified that $\dofbar$ in \eqref{eq:exactDoF} for the high-interference case can be further written as
\begin{align} \label{eq:DoFH}
\dofbar=\begin{cases}
\frac{N}{2}, &D_t\leq 2M-\frac{N}{2}\\
\min\left(\frac{pN+2M-D_t}{2p+1}, \frac{pM}{2p-1}\right), &D_t>2M-\frac{N}{2}
\end{cases}
\end{align}
where $p=\big\lceil\frac{D_t-M}{N-M}\big\rceil$.
In the following, we will separately consider the cases when each term in \eqref{eq:DoFH} is active.

\subsubsection{$\dofbar=\frac{N}{2}$}  With $D_t\leq 2M-\frac{N}{2}$, it can be verified that one feasible choice to \eqref{eq:conditionH} is
\begin{equation}\label{eq:assignNo2}
\begin{aligned}
&|\mathbf{d}_{kh}|=\min\left(M-D_1,\frac{N}{2}\right)\\
&|\mathbf{d}_{km}|=0\\
&|\mathbf{d}_{kt}|=\frac{N}{2}-|\mathbf{d}_{kh}|, \ k=1,2,3.
\end{aligned}
\end{equation}
Since $|\mathbf{d}_{km}|=0$, $\mathbf{E}_{km}$ is an empty matrix. Based on Fig~\ref{F:CoBH}, the total dimension $Z$ occupied by the interference is
\begin{align*}
Z&=|\xtd_{3h}|+|\xtd_{2t}|
=|\mathbf{d}_{3h}|+|\mathbf{d}_{2t}|=\frac{N}{2}.
\end{align*}
As a result, the decodability condition \eqref{eq:lessN} is satisfied with equality; and  hence the DoF value $\dofbar=\frac{N}{2}$ is achievable.

\subsubsection{$\dofbar=\min\big(\frac{pN+2M-D_t}{2p+1}, \frac{pM}{2p-1}\big)$}
In this case, $\dofbar$ can be further written as
\begin{align}\label{eq:DoFHp}
\dofbar=\begin{cases}
\frac{pN+2M-D_t}{2p+1}, &D_t\geq D^{\star}\\
\frac{pM}{2p-1}, &D_t<D^{\star},
\end{cases}
\end{align}
where $D^{\star}\triangleq pN-\frac{2p^2-3p+2}{2p-1}M$.
It can be verified that a feasible choice to \eqref{eq:conditionH} is
\begin{align}
&|\mathbf{d}_{kh}|=M-D_1,\\
&|\mathbf{d}_{km}|=\dofbar-(2M-D_t),\label{eq:dkmH}\\
&|\mathbf{d}_{kt}|=M-D_2.
\end{align}
With such a symbol assignment for each sub-block of $\mathbf d_k$, the total dimension $Z$ occupied by the interference at receiver $1$ can be calculated based on Fig.~\ref{F:CoBH}. Since $\HtdH_{1h}$ and $\HtdH_{1t}$ as defined in \eqref{eq:HkhH} and \eqref{eq:HktH} are full-rank square matrices, interference cancelation can be applied, i.e., the interference caused by $\xtd_{3h}$ and $\xtd_{2t}$ can be subtracted out from $\std_{1m}$ by appropriate linear processing with $\std_{1h}$ and $\std_{1t}$, respectively. Therefore, we have
\begin{align}
Z=&|\xtd_{3h}|+|\xtd_{2t}|+|\mathbf{P}_{12}\xtd_{2m}+\mathbf{P}_{13}\xtd_{3m}|\notag \\
=&|\mathbf{E}_{3h}\mathbf{d}_{3h}|+|\mathbf{E}_{2t}\mathbf{d}_{2t}|+\underbrace{|\mathbf{P}_{12}\mathbf{E}_{2m}\mathbf{d}_{2m}+\mathbf{P}_{13}\mathbf{E}_{3m}\mathbf{d}_{3m}|}_{Z_{m}}\notag \\
=&2M-D_t+Z_{m}, \label{eq:Z1H}
\end{align}
where $\mathbf P_{12}$ and $\mathbf P_{13}$ are full column rank matrices given in \eqref{eq:Pij}.

Next, the middle precoding matrices $\mathbf{E}_{km}\in \mathbb{C}^{(D_t-M)\times |\mathbf d_{km}|}$, with $k=1,2,3$ and $|\mathbf d_{km}|$ given by \eqref{eq:dkmH}, are carefully designed so that the interference dimension $Z_{m}$ is small enough to ensure the decodability condition \eqref{eq:lessN} is satisfied. To this end, interference alignment is required in general. 
 We will illustrate the main idea with $p=1$ and $p=2$, and defer the design of $\mathbf{E}_{km}$ for general $p$ values to Appendix~\ref{A:designEkmGeneralP}.

\textbf{Case I}: $p=1$. By substituting with $p=1$, \eqref{eq:DoFHp} reduces to
\begin{align}\label{eq:achvp1}
\dofbar=\begin{cases}
\frac{N+2M-D_t}{3}, &D_t\geq N-M\\
M, &D_t<N-M
\end{cases}
\end{align}
In this case, the matrices  $\mathbf{E}_{km}$ are randomly generated from $\mathbb{C}^{(D_t-M)\times(\dofbar-2M+D_t)}$. Then we have
 \begin{align}
 Z_{m}=|\mathbf{d}_{2m}|+|\mathbf d_{3m}|=2(\dofbar-2M+D_t).
 \end{align}
 Together with \eqref{eq:Z1H}, we have
 \begin{align}
 \dofbar+Z = 3\dofbar -2M+D_t,
 \end{align}
which can be verified to be no greater than $N$ for both cases in \eqref{eq:achvp1}. Therefore, the full decodability condition \eqref{eq:lessN} is satisfied; and hence the DoF value given in \eqref{eq:achvp1} for $p=1$ is achievable.

\textbf{Case II}: $p=2$. By substituting with $p=2$, \eqref{eq:DoFHp} reduces to
\begin{align}\label{eq:achvp2}
\dofbar=\begin{cases}
\frac{2N+2M-D_t}{5}, &D_t\geq D^{\star}\\
\frac{2M}{3}, &D_t<D^{\star},
\end{cases}
\end{align}
where $D^{\star}=2N-\frac{4M}{3}$.

First, we consider the case $D_t\geq D^{\star}$ so that the objective is to transmit $\dofbar=\frac{2N+2M-D_t}{5}$ data symbols for each user. In this case, we need to apply the technique of ``subspace alignment chains'' introduced in \cite{362} to the middle antennas as shown in Fig.~\ref{F:CoBH} for the design of $\mathbf E_{km}$. 
  Based on \eqref{eq:Pij}, the equivalent system where the subspace alignment chains are applied can be written as $\tilde{\mathbf s}_{km}'=\mathbf P_{k(k+1)} \tilde{\mathbf x}_{(k+1)m}+ \mathbf P_{k(k-1)} \tilde{\mathbf x}_{(k-1)m}$.
As given in \cite{362}, with $\tilde M$ transmit and $\tilde N$ receive antennas, the length of the alignment chain is equal to $\big \lceil \frac{\tilde M}{\tilde N-\tilde M}\big \rceil=p$, where we have $\tilde M=D_t-M$ and $\tilde N=N+D_t-2M$.   Following the notations used in \cite{362}, we use $\mathbf E_{km(s)}^i$, where $k,i\in \{1,2,3\}$, to denote the $s$th $r$-dimensional subspace at transmitter $k$ that participates in the chain originated from transmitter $i$, where $r$ is a design parameter. For the current case considered, we let
\begin{align}
r=\frac{|\mathbf d_{km}|}{2}=\frac{\dofbar-2M+D_t}{2}=\frac{N-4M+2D_t}{5}.
\end{align}



We first consider the alignment chain that originates from transmitter $1$, i.e., $i=1$. With $p=2$,  the chain length is $2$, or equivalently, two subspaces participate the interference alignment for each chain. In this case, the interference alignment must be achieved at receiver $2$, and the alignment chain can be represented as 
\begin{equation}\label{eq:chain1}
\mathbf{E}_{1m(1)}^{1}\overset{\mathrm{Rx}\ 2}{\longleftrightarrow}
\mathbf{E}_{3m(1)}^{1}.
\end{equation}
The  alignment condition \eqref{eq:chain1} is achievable by designing the $r$-dimensional matrices $\mathbf{E}_{1m(1)}^{1}$ and $\mathbf{E}_{3m(1)}^{1}$ such that
\begin{align}\label{eq:alignP2First}
\mathbf P_{21}\mathbf{E}_{1m(1)}^{1}=\mathbf P_{23}\mathbf{E}_{3m(1)}^{1},
\end{align}
\eqref{eq:alignP2First} can be equivalently written as
\begin{equation}\label{eq:alignP2}
\underbrace{\left[\begin{matrix}\mathbf{P}_{21} & -\mathbf{P}_{23}\end{matrix}\right]}_{\mathbf{A}_2}
\underbrace{\left[\begin{matrix}\mathbf{E}_{1m(1)}^{1} \\ \mathbf{E}_{3m(1)}^{1}\end{matrix}\right]}_{\mathbf{G}_2}=\mathbf{0}
\end{equation}

Since $\mathbf{A}_{2}$ is an $(N+D_t-2M)\times 2(D_t-M)$ matrix, the dimension of its null space is $D_t-N$, which can be verified to be no smaller  than $r$ under $p=2$ and $D_t\geq D^{\star}$. Therefore, the columns of $\mathbf{G}_{2}$ can be chosen to be the $r$ basis vectors for the null space of $\mathbf{A}_2$, based on which the matrices $\mathbf{E}_{1m(1)}^{1}$ and $\mathbf{E}_{3m(1)}^{1}$ can be obtained.


With a similar approach, we can design the precoding matrices to achieve two other alignment chains that originate from transmitter 2 and 3, which are represented as
\begin{align}
&\mathbf{E}_{2m(1)}^{2}\overset{\mathrm{Rx}\ 3}{\longleftrightarrow}
\mathbf{E}_{1m(1)}^{2}, \label{eq:chain2} \\
&\mathbf{E}_{3m(1)}^{3}\overset{\mathrm{Rx}\ 1}{\longleftrightarrow}
\mathbf{E}_{2m(1)}^{3}.\label{eq:chain3}
\end{align}

Lastly, the precoding matrices $\mathbf E_{km}\in \mathbb{C}^{(D_t-M)\times |\mathbf d_{km}|}$ for the middle antennas   of the three users are given by
\begin{equation}\label{eq:precoderP2}
\begin{aligned}
&\mathbf{E}_{1m}=\left[\begin{matrix}\mathbf{E}_{1m(1)}^{1} & \mathbf{E}_{1m(1)}^{2}\end{matrix}\right]\\
&\mathbf{E}_{2m}=\left[\begin{matrix}\mathbf{E}_{2m(1)}^{2} & \mathbf{E}_{2m(1)}^{3}\end{matrix}\right]\\
&\mathbf{E}_{3m}=\left[\begin{matrix}\mathbf{E}_{3m(1)}^{1} & \mathbf{E}_{3m(1)}^{3}\end{matrix}\right].
\end{aligned}
\end{equation}
With similar arguments as that in \cite{362}, it can be shown that the precoding vectors obtained above are linearly independent and hence the resulting $\mathbf{E}_{km},k=1,2,3,$ is of full column rank. The interference dimension $Z_{m}$ in \eqref{eq:Z1H} can be calculated as
\begin{align}
Z_{m}=|\mathbf{d}_{2m}|+|\mathbf d_{3m}|-r=\frac{3N-12M+6D_t}{5},
\end{align}
where the minus term is due to the interference alignment achieved by two $r$-dimensional subspaces  at receiver $1$.
Together with \eqref{eq:Z1H} and $\dofbar=\frac{2N+2M-D_t}{5}$, we have
$
\dofbar+Z=N.
$
Therefore, the full decodability condition \eqref{eq:lessN} is satisfied with equality; and hence the DoF given in \eqref{eq:achvp2} for the case $D_t\geq D^{\star}$ is  achievable.

Next, we consider the case  $D_t<D^{\star}$ in \eqref{eq:achvp2} so that the objective is to transmit $\dofbar=\frac{2M}{3}$ data symbols for each user.  Let $r'=D_t-N$ be the dimension of the null space of $\mathbf A_2$ defined in \eqref{eq:alignP2}. First, the $r'$-dimensional precoding matrices are designed similarly to achieve interference alignment given in \eqref{eq:chain1}, \eqref{eq:chain2} and \eqref{eq:chain3}. 
 Since for each user, two such precoding matrices are used as shown in \eqref{eq:precoderP2}, the number of data symbols participating interference alignment for each user is given by $2r'=2(D_t-N)$, which is less than $|\mathbf d_{km}|=\frac{2M}{3}-2M+D_t$ as required by \eqref{eq:dkmH} under $D_t<D^{\star}$. Therefore, for each user $k$, we need to transmit another $\hat r$ data symbols with precoding matrices denoted as $\hat{\mathbf E}_{km}$, where
\begin{align*}
\hat r=&|\mathbf{d}_{km}|-2r'=\frac{2M}{3}-(2M-D_t)-2(D_t-N)\\
=&\big(2N-\frac{4M}{3}\big)-D_t.
\end{align*}
Furthermore, $\hat{\mathbf E}_{km}$ are randomly generated. Then the precoding matrices $\mathbf E_{km}$ for the middle antennas are obtained as
\begin{align*}
&\mathbf{E}_{1m}=\left[\begin{matrix}\mathbf{E}_{1m(1)}^{1} & \mathbf{E}_{1m(1)}^{2} & \hat{\mathbf E}_{1m}\end{matrix}\right]
\end{align*}
\begin{align*}
&\mathbf{E}_{2m}=\left[\begin{matrix}\mathbf{E}_{2m(1)}^{2} & \mathbf{E}_{2m(1)}^{3} & \hat{\mathbf E}_{2m}\end{matrix}\right]
\end{align*}
\begin{align*}
&\mathbf{E}_{3m}=\left[\begin{matrix}\mathbf{E}_{3m(1)}^{1} & \mathbf{E}_{3m(1)}^{3} & \hat{\mathbf E}_{3m}\end{matrix}\right]
\end{align*}
It can be verified that the resulting $\mathbf{E}_{km}$ is of full column rank. The interference dimension $Z_{m}$ at the middle antennas of receiver $1$ is given by
\begin{align*}
Z_{m}=&|\mathbf d_{2m}|+|\mathbf d_{3m}|-r'
=D_t+N-\frac{8M}{3}.
\end{align*}
Together with \eqref{eq:Z1H}, we have
\begin{align*}
\dofbar+Z=\frac{2M}{3}+(2M-D_t)+\big(D_t+N-\frac{8M}{3}\big)=N.
\end{align*}
Therefore, the decodability condition \eqref{eq:lessN} is satisfied with equality; and hence $\dofbar=\frac{2M}{3}$ is achievable.

\section{Numerical Results}\label{sec:numerical}
In this section, we provide a numerical example for the proposed  techniques. 
We assume that each terminal is equipped with a uniform linear array (ULA) with adjacent elements separated by distance $\Delta$, where $\Delta$ is measured in wavelength. The channel matrix $\mathbf H_{ki}$ between transmitter $i$ and receiver $k$ can then be modeled as \cite{474}
\begin{align}\label{eq:HULA}
\mathbf H_{ki}=\frac{1}{\sqrt{L_{ki}}}\sum_{l=1}^{L_{ki}} \mathbf a_R(\phi_{ki}^l)\mathbf a_T^H(\theta_{ki}^l),\ k,i=1,2,3,
\end{align}
where $L_{ki}$ represents the number of signal paths; $\phi_{ki}^l$ and $\theta_{ki}^l$  denote the angle of arrival (AoA) and angle of departure (AoD) for the $l$th path between transmitter $i$ and receiver $k$, respectively. Moreover, $\mathbf a_T$ and $\mathbf a_R$ are the transmit and receive array response, respectively, which are given by
\begin{align}
\mathbf a_T(\theta)&=\left[\begin{matrix}1, e^{j2\pi \Delta \sin \theta}, \cdots, e^{j2\pi\Delta(M_T-1)\sin \theta} \end{matrix}\right]^T,\\
\mathbf a_R(\phi)&=\left[\begin{matrix}1, e^{j2\pi \Delta \sin \phi}, \cdots, e^{j2\pi\Delta(M_R-1)\sin \phi} \end{matrix}\right]^T.
\end{align}
Since $\mathbf H_{ki}$ in \eqref{eq:HULA} is  given by a summation of $L_{ki}$ rank-1 matrices, it is clear that in the poor scattering environment with $L_{ki}<\min(M_T,M_R)$, the channel matrix $\mathbf H_{ki}$ will be rank deficient.

We consider a 3-user MIMO-IC with $M_T=2$ and $M_R=4$, and all direct links have two paths $L_{kk}=2$ and the interfering links have one single path only $L_{ki}=1, \forall k\neq i$. With the AoAs and AoDs randomly generated, this corresponds to a rank-deficient MIMO-IC with $D_0=2$ and $D_1=D_2=1$. Based on Theorem~\ref{theo:main}, the DoF per user is $\bar d=2$, i.e., each receiver is expected to be able to detect two data streams sent by their respective transmitters. As a numerical illustration, we consider a particular channel realization for a set of  randomly generated AoAs and AoDs given by
\begin{align}
\boldsymbol \Phi^1 =\left[\begin{matrix}
0.05  &   3.47    &  2.47 \\
    4.57    &  4.66    &  0.85 \\
    3.53    &  5.18   &   1.23
\end{matrix}\right], \quad
\boldsymbol \Theta^1 =\left[\begin{matrix}
 4.53    &  2.41    &  0.93 \\
    4.21   &   3.09    &  3.05 \\
    0.38    &  4.86    &  0.45
\end{matrix}\right]
\end{align}
\begin{align}
\boldsymbol \phi^2=\left[\begin{matrix}4.95  &   2.47 &    1.48  \end{matrix}\right], \quad
\boldsymbol \theta^2=\left[\begin{matrix} 0.51   &  0.50  &  5.83  \end{matrix}\right],
\end{align}
where the $(k,i)$th entry of $\boldsymbol \Phi^1$ and $\boldsymbol \Theta^1$ represent the AoA and AoD of the first path for the link between transmitter $i$ and receiver $k$, respectively, and the $k$th entry of $\boldsymbol \phi^2$ and $\boldsymbol \theta^2$ are the AoA and AoD of the second path for the direct link $\mathbf H_{kk}$. With the scheme presented in Section~\ref{sec:innerBound}, the precoding matrices $\mathbf E_k$ in Fig.~\ref{F:precoding} can be any randomly generated $2\times 2$ diagonal matrices, and $\mathbf T_k$ can be found as
\begin{align}
\mathbf T_1&=\left[\begin{matrix}
 0.37 - 0.33i & -0.48  + 0.13i\\
  -0.45 - 0.21i & -0.50 - 0.07i
\end{matrix}\right],\
  \mathbf T_2=\left[\begin{matrix}
  -0.40 - 0.31i & -0.50 + 0.01i\\
  -0.46 + 0.18i & -0.50 - 0.01i
  \end{matrix}\right], \\
  \mathbf T_3&=\left[\begin{matrix}
   0.49 - 0.10i & -0.46 - 0.20i\\
  -0.50 - 0.05i & -0.49 + 0.11i
  \end{matrix}\right],
\end{align}

Furthermore, with the following  linear transformation applied at receiver $1$:
\begin{align}
\mathbf R_1=\left[\begin{matrix}
 -0.34 - 0.08i &  0.11 + 0.34i  & 0.31 - 0.18i &  -0.31 - 0.16i\\
  -0.46       &     -0.02 + 0.54i & -0.30 + 0.28i  & 0.41 + 0.41i\\
   0.55      &      -0.02 + 0.45i &  0.28 - 0.50i  & 0.37 + 0.19i\\
   0.29 + 0.20i & -0.08 + 0.35i & -0.31 + 0.18i & -0.27 - 0.22i
 \end{matrix}\right],
\end{align}
it can be verified that the second and the third antennas of $\tilde{\mathbf y}_1$ only contain the two desired data symbols sent from transmitter $1$, and hence can be used to fully decode the desired symbols. Similarly, receiver $2$ and $3$ can both detect their desired data symbols after applying $\mathbf R_2$ and $\mathbf R_3$, respectively.

\section{Conclusion}\label{sec:conclusion}
In this paper, we have provided the complete DoF characterization for the spatially-normalized DoF of the $3$-user rank-deficient MIMO-IC parameterized by $(M_T, M_R, D_0, D_1, D_2)$. By exploiting the rank-deficiency of the interfering channel matrices and applying invertible linear transformations, we first convert the original fully-connected rank-deficient MIMO-IC into an equivalent partially-connected full-rank MIMO-IC, based on which the existing techniques developed for full-rank channels can be incorporated to facilitate the outer and inner bounds derivations. The outer bound is obtained by the genie-aided signaling technique, where the appropriate genie signals to be provided to the receivers are determined based on the developed equivalent channel model. To achieve the optimal DoF,  a two-layered linear processing scheme that combines zero forcing, interference alignment and interference cancelation is proposed. Therefore, for rank-deficient MIMO-ICs, zero forcing or interference alignment alone is not sufficient to achieve the optimal DoF; instead, a combination of the two is required in general.

While we have shown the DoF achievability using the technique of spatial extension for non-integer DoF values, whether the outer bound can still be achieved if we restrict to time/frequency extension only remains an open problem. The additional complexity introduced by such a restriction is the resulting block diagonal channel structure, which needs to be handled more carefully. Another open problem is how to extend the techniques developed in this paper to the completely asymmetric rank-deficient scenario. This will be a challenging task since the design of the zero-forcing matrices, for instance, is much more involved due to a considerably larger number of parameters need to be considered.


\appendices

\section{Achievability of $\dofbar=\min\big(\frac{pN+2M-D_t}{2p+1}, \frac{pM}{2p-1}\big)$}\label{A:designEkmGeneralP}
In this section, we show the achievability of $\dofbar=\min\big(\frac{pN+2M-D_t}{2p+1}, \frac{pM}{2p-1}\big)$  in \eqref{eq:DoFHp} for general $p$.
\subsection{Achievability of $\dofbar=\frac{pN+2M-D_t}{2p+1}$}
 We first consider the case when $D_t\geq D^{\star}$ so that $\dofbar=\frac{pN+2M-D_t}{2p+1}$. Then from \eqref{eq:dkmH}, we have $|\mathbf d_{km}|=\frac{p(N-4M+2D_t)}{2p+1}$, which is the number of independent symbols need to be transmitted by each user via the middle $D_t-M$ antennas with precoding matrix $\mathbf E_{km}$. With  subspace alignment chains technique, we use $\mathbf E_{km(s)}^i$, where $k,i\in \{1,2,3\}$, to denote the $s$th $r$-dimensional subspace at transmitter $k$ that participates in the chain originated from transmitter $i$. Here, we let $r=\frac{|\mathbf d_{km}|}{p}=\frac{N-4M+2D_t}{2p+1}$.

We first consider the alignment chain that originates from transmitter $1$. With $p=\big\lceil\frac{\Mtd}{\Ntd-\Mtd}\big\rceil$, where $\Mtd=D_t-M$ and $\Ntd=N+D_t-2M$ representing the number of transmit and receive antennas in the middle block as shown in Fig.~\ref{F:CoBH}, it follows from \cite{362} that the chain length is equal to $p$.
Following similar notations used in \cite{362}, we let $\underline{p}=(p\mod 3)$, $\overline{p}=\lceil \frac{p}{3}\rceil$, $a(1)=0$, $a(2)=3$ and $a(0)=2$. Then the  alignment chain that originates from transmitter $1$ can be represented as
\begin{align}
&\mathbf{E}_{1m(1)}^{1}\overset{\mathrm{Rx}\ 2}{\longleftrightarrow}
\mathbf{E}_{3m(1)}^{1}\overset{\mathrm{Rx}\ 1}{\longleftrightarrow}
\mathbf{E}_{2m(1)}^{1}\overset{\mathrm{Rx}\ 3}{\longleftrightarrow}
\mathbf{E}_{1m(2)}^{1}\overset{\mathrm{Rx}\ 2}{\longleftrightarrow}
\mathbf{E}_{3m(2)}^{1}\cdots \mathbf{E}_{a(\underline{p-1})m(\overline{p-1})}^{1}
\overset{\mathrm{Rx}\ a(\underline{p+1})}{\longleftrightarrow}
\mathbf{E}_{a(\underline{p})m({\overline{p}})}^{1}.\label{eq:Nchain1}
\end{align}

The alignment chain in \eqref{eq:Nchain1} can be  equivalently written as
$\mathbf{A}\mathbf{G}=\mathbf{0}$,
where
\begin{equation*}
\footnotesize
\mathbf{A}=\left[
\begin{matrix}
\mathbf{P}_{21} & -\mathbf{P}_{23} & \mathbf{0} & \mathbf{0} & \mathbf{0} & \mathbf{0}\\
\mathbf{0} & \mathbf{P}_{13} & -\mathbf{P}_{12} & \mathbf{0} & \mathbf{0} & \mathbf{0}\\
\mathbf{0} & \mathbf{0} & \mathbf{P}_{32} & -\mathbf{P}_{31} & \mathbf{0} & \mathbf{0}\\
\mathbf{0} & \mathbf{0} & \ddots & \ddots & \vdots & \mathbf{0}\\
\mathbf{0} & \mathbf{0} & \cdots & \mathbf{0} & \mathbf{P}_{a(\underline{p+1})a(\underline{p-1})} & -\mathbf{P}_{a(\underline{p+1})a(\underline{p})}
\end{matrix}
\right],
\
\mathbf{G}=\left[\begin{matrix}
\mathbf{E}_{1m(1)}^{1} \\
\mathbf{E}_{3m(1)}^{1} \\
\vdots\\
\mathbf{E}_{a(\underline{p})m({\overline{p}})}^{1}
\end{matrix}\right].
\end{equation*}

Since $\mathbf{A}$ is a $(p-1)\Ntd\times p\Mtd$ matrix, the dimension of its null space is given by $\mathrm{dim}(\mathcal{N}(\mathbf{A}))=p\Mtd-(p-1)\Ntd=D_t-(p-1)N+(p-2)M$. With $D_t\geq D^{\star}$, it can be verified that $r\leq\mathrm{dim}(\mathcal{N}(\mathbf{A}))$ is always satisfied. Therefore, the $r$ linearly independent columns of $\mathbf{G}$ are chosen from the null space of $\mathbf A$, from which we can obtain $\mathbf{E}_{1m(1)}^{1},\cdots, \mathbf{E}_{a(\underline{p})m({\overline{p}})}^{1}$.

Similarly, we can design the precoding matrices for two other $r$-dimensional subspace alignment chains with chain length $p$  that originate from transmitter $2$ and transmitter $3$, respectively. The precoding matrix $\mathbf E_{km}$ for the middle antennas is then given by
\begin{align*}
\mathbf{E}_{km}=\big[
& \mathbf{E}_{km(1)}^{1},\mathbf{E}_{km(2)}^{1},\cdots, \mathbf{E}_{km(1)}^{2}, \mathbf{E}_{km(2)}^{2},\cdots, \mathbf{E}_{km(1)}^{3}, \mathbf{E}_{km(2)}^{3}, \cdots
\big].
\end{align*}

Since there are $pr$ vectors participating in each alignment chain, in total $3pr$ symbols are transmitted  from the middle antennas by all the three users. Due to symmetry, each user transmits $pr$ symbols by the the middle antennas, which is equal to $|\mathbf d_{km}|$, as desired.
Following similar arguments as that in \cite{362}, it can be shown that the columns in $\mathbf{E}_{km}$ are linearly independent and $\mathbf{E}_{km}$ is full column rank.

With $r$-dimensional subspace alignment chain of length $p$, it follows that $(p-1)r$-dimensional interfering signals are aligned at receiver $1$. Therefore, we have
\begin{align*}
Z_{m}&=|\mathbf{d}_{2m}|+|\mathbf{d}_{3m}|-(p-1)r=(p+1)r=\frac{(p+1)(N-4M+2D_t)}{2p+1}.
\end{align*}
Together with \eqref{eq:Z1H} and $\dofbar=\frac{pN+2M-D_t}{2p+1}$, we have
$
\dofbar+Z=N.
$
Therefore, the full decodability condition \eqref{eq:lessN} is satisfied with equality, hence $\dofbar=\frac{pN+2M-D_t}{2p+1}$ is achievable.

\subsection{Achievability of $\dofbar=\frac{pM}{2p-1}$}
 With $D_t<D^{\star}$, we show that each user is able to transmit $\frac{pM}{2p-1}$ independent data streams.  From \eqref{eq:dkmH}, the number of symbols $|\mathbf d_{km}|$ need to be transmitted by the $D_t-M$ middle antennas is given by $|\mathbf d_{km}|=D_t-\frac{(3p-2)M}{2p-1}$. To this end, we need to use two groups of alignment chains, one with chain length $p$ and subspace dimension $r'$, and the other with chain length $p-1$ and subspace dimension $\hat r$.  The design of the first group of alignment chains follows exactly the same manner as the previous case, except that now the number of columns of $\mathbf{G}$ is set to be $r'=\mathrm{dim}(\mathcal{N}(\mathbf{A}))=D_t-(p-1)N+(p-2)M$. The number of symbols transmitted by the first group of alignment chains from the middle antennas is then given by $pr'$, which is less than $|\mathbf d_{km}|$ under the condition $D_t<D^{\star}$.  Therefore, we need to use another group of alignment chains of length $p-1$. For this group of alignment chains, we use $\widehat {\mathbf E}_{km(s)}^i$, where $k,i\in \{1,2,3\}$, to denote the $s$th $\hat r$-dimensional subspace at transmitter $k$ that participates in the chain originated from transmitter $i$. To ensure $|\mathbf d_{km}|$ data symbols sent by the middle antennas,  we let
\begin{align*}
\hat r &=\frac{|\mathbf{d}_{km}|-r'p}{p-1}=D_t-D^{\star}.
\end{align*}

 Then the alignment chain originated from transmitter $1$ can be represented as
\begin{equation}\label{eq:NchainGroup2}
\begin{aligned}
&\widehat {\mathbf E}_{1m(1)}^{1}\overset{\mathrm{Rx}\ 2}{\longleftrightarrow}
\widehat {\mathbf E}_{3m(1)}^{1}\overset{\mathrm{Rx}\ 1}{\longleftrightarrow}
\widehat {\mathbf E}_{2m(1)}^{1}\overset{\mathrm{Rx}\ 3}{\longleftrightarrow}
\widehat {\mathbf E}_{1m(2)}^{1}\cdots \widehat {\mathbf E}_{a(\underline{p-2})m(\overline{p-2})}^{1}
\overset{\mathrm{Rx}\ a(\underline{p})}{\longleftrightarrow}
\widehat {\mathbf E}_{a(\underline{p-1})m({\overline{p-1}})}^{1},
\end{aligned}
\end{equation}
or equivalently
\begin{align}
\widehat{\mathbf{A}}\widehat{\mathbf{G}}=\mathbf{0},\label{eq:NchainGroup2Null}
\end{align}
where $\widehat{\mathbf{A}}$ is a sub-matrix of $\mathbf{A}$ given by
\begin{equation*}
\footnotesize
\widehat{\mathbf{A}}=\left[
\begin{matrix}
\mathbf{P}_{21} & -\mathbf{P}_{23} & \mathbf{0} & \mathbf{0} & \mathbf{0} & \mathbf{0}\\
\mathbf{0} & \mathbf{P}_{13} & -\mathbf{P}_{12} & \mathbf{0} & \mathbf{0} & \mathbf{0}\\
\mathbf{0} & \mathbf{0} & \mathbf{P}_{32} & -\mathbf{P}_{31} & \mathbf{0} & \mathbf{0}\\
\mathbf{0} & \mathbf{0} & \ddots & \ddots & \vdots & \mathbf{0}\\
\mathbf{0} & \mathbf{0} & \cdots & \mathbf{0} & \mathbf{P}_{a(\underline{p})a(\underline{p-2})} & -\mathbf{P}_{a(\underline{p})a(\underline{p-1})}
\end{matrix}
\right],\
\widehat{\mathbf{G}}=\left[\begin{matrix}
\widehat{\mathbf{E}}_{1m(1)}^{1} \\
\widehat{\mathbf{E}}_{3m(1)}^{1} \\
\vdots\\
\widehat{\mathbf{E}}_{a(\underline{p-1})m({\overline{p-1}})}^{1}
\end{matrix}\right].
\end{equation*}

The size of $\widehat{\mathbf{A}}$ is $(p-2)\Ntd\times(p-1)\Mtd$ and thus the dimension of its null space is given by  $\mathrm{dim}(\mathcal{N}(\widehat{\mathbf{A}}))=(p-1)\Mtd-(p-2)\Ntd=D_t-(p-2)N+(p-3)M$. Note that since the submatrix of $\mathbf G$ formed by its first $(p-1)\Mtd$ rows also satisfy \eqref{eq:NchainGroup2Null}. To ensure the linear independence among the precoding vectors, the $\hat r$ columns of $\widehat{\mathbf G}$ are selected from the null space of $\widehat{\mathbf{A}}$, but not in the subspace spanned by the columns of the aforementioned submatrix of $\mathbf G$. The dimension is
\begin{align*}
\mathrm{dim}(\mathcal{N}(\widehat{\mathbf{A}}))-\mathrm{dim}(\mathcal{N}(\mathbf{A}))=N-M.
\end{align*}
It can be verified that  $\hat r\leq N-M$, therefore, the $\hat r$ columns of $\widehat {\mathbf G}$ can be obtained. Similarly, the precoding vectors $\widehat {\mathbf E}_{k(s)}^2$ and  $\widehat {\mathbf E}_{k(s)}^3$ for the other two alignment chains originating from transmitter $2$ and $3$ can be found. The overall precoding matrix $\mathbf E_k$ for the middle antennas of user $k$ is then obtained accordingly. For the two groups of alignment chains discussed above, the number of aligned interference dimensions at each receiver are respectively given by $r'(p-1)$ and $\hat r(p-2)$. Therefore, with the proposed precoder design, we have
\begin{align}
Z_{m}=
&|\mathbf{d}_{2m}|+|\mathbf{d}_{3m}|-r'(p-1)-\hat r(p-2)\notag\\
=&N-(2M-D_t)-\frac{pM}{2p-1}.
\end{align}
Then together with $\eqref{eq:Z1H}$, we have $\dofbar+Z=N$. In other words, the full decodability condition \eqref{eq:lessN} is satisfied with equality. Therefore, $\dofbar=\frac{pM}{2p-1}$ is achievable.

\bibliographystyle{IEEEtran}
\bibliography{IEEEabrv,IEEEfull}
\end{document}